\documentclass[aip,jcp,reprint,superscriptaddress,groupedaddress,superscriptaddress]{revtex4-2}
\usepackage[english]{babel}
\usepackage[LGR,T1]{fontenc}
\usepackage{bigstrut}
\usepackage{graphicx}
\usepackage{dcolumn}
\usepackage{bm}
\usepackage{amsmath}
\usepackage{amsfonts,amssymb}
\usepackage[colorlinks=true,linkcolor=teal,citecolor=teal,urlcolor=teal]{hyperref}
\usepackage[shortlabels]{enumitem}
\usepackage{xcolor}
\usepackage{nomencl}
\usepackage{comment}
\usepackage[version=4]{mhchem}
\usepackage{orcidlink}
\usepackage{phonetic}
\usepackage{multirow}
\usepackage{natbib}
\makenomenclature

\begin{document}

\title{QMeCha: quantum Monte Carlo package for fermions in embedding environments}  

\author{Matteo Barborini\orcidlink{0000-0001-7798-099X}}
\email[Corresponding Author: ]{matteo.barborini@gmail.com}
\affiliation{HPC Platform, University of Luxembourg, L-4364 Esch-sur-Alzette, Luxembourg\looseness=-3}
\affiliation{Department of Physics and Materials Science, University of Luxembourg, L-1511 Luxembourg City, Luxembourg \looseness=-3}

\author{Jorge Alfonso Charry Martínez\orcidlink{0000-0003-3069-2522}}
\affiliation{Department of Physics and Materials Science, University of Luxembourg, L-1511 Luxembourg City, Luxembourg \looseness=-3}

\author{Matej Ditte\orcidlink{0009-0001-8617-1358}}
\affiliation{Department of Physics and Materials Science, University of Luxembourg, L-1511 Luxembourg City, Luxembourg \looseness=-3}

\author{Andronikos Leventis\orcidlink{0009-0002-3097-0617}}
\affiliation{Department of Physics and Materials Science, University of Luxembourg, L-1511 Luxembourg City, Luxembourg \looseness=-3}

\author{Georgios Kafanas\orcidlink{0000-0002-9048-5296}} 
\affiliation{HPC Platform, University of Luxembourg, L-4364 Esch-sur-Alzette, Luxembourg\looseness=-3}

\author{Alexandre Tkatchenko\orcidlink{0000-0002-1012-4854}}
\affiliation{Department of Physics and Materials Science, University of Luxembourg, L-1511 Luxembourg City, Luxembourg \looseness=-3}

\date{\today}

\begin{abstract}
We present the first \textit{open access} version of the {\sffamily QMeCha} (Quantum MeCha /'m\epsi k\schwa /) code, a quantum Monte Carlo (QMC) package developed to study many-body interactions between different types of quantum particles, with a modular and easy-to-expand structure.
{\sffamily QMeCha} is now available under CC BY-NC-ND license through the repository {\ttfamily github.com/QMeCha}.
The present code has been built to solve the Hamiltonian of a system that can include nuclei and fermions of different mass and charge, e.g. electrons and positrons, embedded in an environment of classical charges and quantum Drude oscillators. 
To approximate the ground state of this many-particle operator, the code features different wavefunctions. 
For the fermionic particles, beyond the traditional Slater determinant, {\sffamily QMeCha} also includes Geminal functions such as the Pfaffian, and presents different types of explicit correlation terms in the Jastrow factors.
The classical point charges and quantum Drude oscillators, described through different variational ans\"atze, are used to model a molecular environment capable of explicitly describing dispersion, polarization, and electrostatic effects experienced by the nuclear and fermionic subsystem. 
To integrate these wavefunctions, efficient variational Monte Carlo and diffusion Monte Carlo protocols have been developed, together with a robust wavefunction optimization procedure that features correlated sampling.
%
\end{abstract}

\maketitle

\tableofcontents

\nomenclature[01]{$N_n$}{Number of nuclei}
\nomenclature[02]{$N_e$}{Number of electrons}
\nomenclature[03]{$N_p$}{Number of positrons}
\nomenclature[04]{$N_f$}{Number of electrons and positrons in the system ($N_e+N_p$)}
\nomenclature[05]{$N_O$}{Number of Quantum Drude Oscillators}
\nomenclature[06]{$N_C$}{Number of fixed point-charges}
\nomenclature[07]{$\textbf{r}$}{Coordinate vector of a single electron or positron}
\nomenclature[08]{$\sigma$}{Spin state of a single electron or positron}
\nomenclature[09]{$\textbf{r}^e$}{$3$-dimensional vector of an electron coordinate}
\nomenclature[10]{$\textbf{r}^p$}{$3$-dimensional vector of a positron coordinate}
\nomenclature[11]{$\textbf{r}^d$}{$3$-dimensional vector of a drudonic coordinate}
\nomenclature[12]{$\bar{\textbf{r}}^d$}{$3N_O$-dimensional vector of the drudonic coordinates $\bar{\textbf{r}}^d=\{\textbf{r}^d_i\}_{i=1}^{N_O}$}
\nomenclature[13]{$\bar{\textbf{r}}^e$}{$3N_e$-dimensional vector of the electronic coordinates $\bar{\textbf{r}}^e=\{\textbf{r}^e_i\}_{i=1}^{N_e}$}
\nomenclature[14]{$\bar{\textbf{r}}^p$}{$3N_p$-dimensional vector of the positronic coordinates $\bar{\textbf{r}}^p=\{\textbf{r}^p_i\}_{i=1}^{N_p}$}
\nomenclature[15]{$\bar{\textbf{r}}^f$}{$3N_f$-dimensional vector of electronic and positronic coordinates $\bar{\textbf{r}}^f=\{\textbf{r}_i\}_{i=1}^{N_f}$}
\nomenclature[16]{$\bar{\textbf{r}}$}{$3N$-dimensional vector of the Cartesian coordinates of all particles in the system electrons, positrons and drudons (if present) $\bar{\textbf{r}}=\{\textbf{r}_i\}_{i=1}^{N}$}
\nomenclature[17]{$m_i,q_i$}{Mass and charge of the $i$th fermion (electron or positron).}
\nomenclature[17]{$M_a,Z_a$}{Mass and charge of the $a$th nucleus.}
\nomenclature[18]{$\mu_i,\rho_i,\omega_i,\textbf{R}_i^O$}{Mass, charge, frequency, and Cartesian vector of the position of the centre of the $i$th quantum Drude Oscillator.}
\nomenclature[18]{$\textbf{R}_a$}{Vector of the position of the $a$th nuclei.}
\nomenclature[18]{$\bar{\textbf{R}}$}{$3N_n$-dimensional vector of the nuclear coordinates $\bar{\textbf{R}}=\{\textbf{R}_a\}_{a=1}^{N_n}$}
\nomenclature[19]{$\textbf{R}^C_i$}{Vector of the position of the $i$th point charge in the embedding environment.}
\nomenclature[20]{$Q_i$}{Value of the $i$th point charge in the embedding environment.}
\nomenclature[21]{$ \hat{\text{H}}_{tot}$}{Total Hamiltonian including molecular systems and embedding environment.}
\nomenclature[22]{$ \hat{\text{H}}_{f}$}{Molecular Hamiltonian, including positrons if present.}
\nomenclature[23]{$ \hat{\text{H}}_{O}$}{Hamiltonian of the system of QDOs.}
\nomenclature[24]{$ \nabla^2_{\textbf{r}} $}{Laplacian operator for the general 3-coordinates vector $\textbf{r}$.}
\nomenclature[25]{$ \text{V}(\bar{\textbf{r}};\bar{\textbf{R}})$}{Potential energy including all the Coulomb interactions between electrons, positrons and nuclei.}
\nomenclature[26]{$\hat{h}_{i}^{O}$}{Single particle Hamiltonian for the $i$th drudon.}
\nomenclature[27]{$\text{V}_{O,f}$}{Coulomb interactions between molecular system and the QDOs.}
\nomenclature[28]{$\text{V}_{C,f}$}{Coulomb interactions between molecular system and the point charges in the embedding environment.}
\nomenclature[29]{$\hat{\text{V}}^a_{\text{ECP}}(\textbf{r})$}{Effective core potential for the $a$th nuclei.}
\nomenclature[30]{$\text{V}^a_{loc}(\textbf{r})$ }{Local part of the effective core potential for the $a$th nuclei.}
\nomenclature[31]{$\text{V}^a_{l}(\textbf{r})$ }{Non-local part of the effective core potential for the $a$th nuclei with angular momentum $l$.}
\nomenclature[32]{$\hat{\text{P}}^a_{l}$ }{Projection operator on the real spherical harmonic centred on the $a$th nuclei with angular momentum $l$.}
\nomenclature[33]{$\textbf{E}$}{Vector of the external static electric field.}
\nomenclature[34]{$\boldsymbol{\mu}$}{Vector of the total dipole of the system}
\nomenclature[35]{$\text{V}_{\textbf{E}}$ }{Energy contribution from the external electric field $\textbf{E}$.}
\nomenclature[36]{$\Psi_T(\bar{\textbf{x}},\bar{\textbf{r}}^d)$ }{Total trial wavefunction of the molecular system and the embedding environment.}
\nomenclature[37]{$\Psi_f(\bar{\textbf{x}}) $ }{Trial wavefunction of the molecular system including eventually the positronic coordinates.}
\nomenclature[38]{$\Psi_d(\bar{\textbf{r}}^d)$ }{Wavefunction of the drudons in the embedding environment.}
\nomenclature[39]{$\Psi_{f,d}(\bar{\textbf{r}},\bar{\textbf{r}}^d)$ }{Correlation function between the molecular system and the drudons in the environment.}
\nomenclature[40]{$\psi_e(\bar{\textbf{x}}^e)$ }{Electronic part of the wavefunction, only depends on electronic and parametric nuclear coordinates.}
\nomenclature[41]{$\psi_p(\bar{\textbf{x}}^p)$ }{Positronic part of the wavefunction, only depends on positronic and parametric nuclear coordinates.}
\nomenclature[42]{$\mathcal{J}(\bar{\textbf{x}})$ }{Total Jastrow factor.}
\nomenclature[43]{$J_c(\bar{\textbf{x}})$ }{Cusp terms in the Jastrow factor.}
\nomenclature[43]{$J_d(\bar{\textbf{r}})$ }{Jastrow factor including the 3/4 body terms.}
\nomenclature[43]{$\text{pf}[\textbf{P}]$ }{Pfaffian of the matrix $\textbf{P}$}
\nomenclature[44]{$\textbf{G}$ }{Singlet geminal matrix.}
\nomenclature[45]{$\textbf{T}^{\uparrow}$,$\textbf{T}^{\downarrow}$ }{Triplet geminal matrices, for spin up and spin down particles of the same type.}
\nomenclature[46]{$\textbf{G}_{ij} = \phi_G ( \textbf{r}^{\uparrow}_i, \textbf{r}^{\downarrow}_j )$}{Singlet geminal coupling for the $i$th and $j$th electrons of opposite spin.}
\nomenclature[47]{$\textbf{T}^{\sigma }_{ij} = \phi_T ( \textbf{r}^{\sigma}_i, \textbf{r}^{\sigma}_j )$ }{Triplet geminal coupling for the $i$th and $j$th electrons of identical spin ($\sigma = \{\uparrow,\downarrow\}$).}
\nomenclature[48]{$\textbf{S}^{\sigma}$ }{Slater matrices for the two spin populations ($\sigma = \{\uparrow,\downarrow\}$).}

\nomenclature[51]{$\text{E}\left[\Psi_T(\boldsymbol{\alpha})\right]$ }{Energy functional on the trial wavefunction $\Psi_T(\boldsymbol{\alpha})$.}
\nomenclature[52]{$\boldsymbol{\alpha}$ }{Parameters' vector of the trial wavefunction.}
\nomenclature[53]{$\text{E}_{l}(\bar{\textbf{r}}; \boldsymbol{\alpha})$ }{Local energy.}
\nomenclature[54]{$\Pi(\bar{\textbf{r}}; \boldsymbol{\alpha})$ }{Probability density}
\nomenclature[55]{$\bar{\text{E}}_{l}(\boldsymbol{\alpha})$ }{Sample mean value of the local energy.}
\nomenclature[56]{$s^2_{\bar{\text{E}}_l(\boldsymbol{\alpha})}$ }{Sample variance of the local energy distribution.}
\nomenclature[57]{$\sigma_{\bar{\text{E}}_l(\boldsymbol{\alpha})}$ }{Statistical error on the mean of the local energy.}
\nomenclature[58]{$f_{\alpha_k}$ }{Generalized force for parameter $\alpha$. Derivative of the energy functional with respect to $\alpha_k$.}
\nomenclature[59]{$\mathcal{O}_{\alpha_k}$ }{Derivative of the logarithm of the wavefunction with respect to the parameter $\alpha_k$.}
\nomenclature[60]{$\textbf{S}$}{Covariance matrix of the local operators $\mathcal{O}_{\alpha_k}$ on the sampling.}

\printnomenclature
\markboth{}{}

\section{Introduction}\label{sec:intro}

The solution of the many-body Schr{\"o}dinger equation~\cite{sch+26pr} for molecular or periodic systems of atomic nuclei and electrons, has been one of the main challenges for theoretical physicists and chemists since its conceptualization.~\cite{Heitler1927,Hartree1928,Slater1928,Fock1930,Slater1930,koopmans34,low+50jcp,roo+51rmp,rue+54jcp}.
The daunting task of constructing an analytical solution to the problem of interacting nuclei and electrons led to the development of numerous models to tackle its solution at least in an approximate manner.\cite{Heitler1927,Hartree1928,Slater1928,Fock1930,Slater1930,koopmans34,low+50jcp}
A substantial breakthrough occurred with the use of the first efficient computers in the mid-50s\cite{roo+51rmp,rue+54jcp,pop+55rsa,sch+55jcp,boys+56nat} that stimulated the development of a wide range of progressively more accurate numerical methods.

The most successful of these computational approaches fall into three main categories. These include wavefunction-based methods, such as Hartree-Fock (HF)\cite{Hartree1928,Slater1928,Fock1930,Slater1930}, Configuration Interaction\cite{Craig50,Roos52,Sherrill+99} (CI) and Coupled Cluster\cite{coe+66np,cizek+66jcp,paldus+72pra,bartlett+07rmp} (CC) multi-determinantal expansions, complete active-space (CAS) theories\cite{Roos+80cp}; the long list of perturbative approaches like M{\o}ller-Plesset (MP) perturbation theory\cite{MP34} and symmetry-adapted perturbation theory (SAPT)\cite{SAPT94}; and the density-based methods, such as Density Functional Theory (DFT)\cite{HohenbergKohn64,KohnSham65}.

Within the framework of wavefunction-based approaches, quantum Monte Carlo (QMC)\cite{mcm+65pr,cep+77prb,fou+01rmp} methods have established themselves as efficient and versatile alternatives, due to two main features. 
First, the stochastic nature of their algorithms renders them highly parallelizable on modern high-performance computing (HPC) facilities. This together with the favourable scaling of the computational cost with respect to the number of particles $N$ in the system, $\mathcal{O}(N^{3\sim4})$, brings a competitive advantage for large systems when compared to the previously discussed deterministic approaches.
Second, the versatility of the QMC algorithms also allows to work with more sophisticated many-body wavefunctions that can explicitly include correlation between particles \cite{bec+17}, compared to single-particle orbitals in HF-based methods.  
In fact, in the first article, published in 1977, in which Monte Carlo methods were applied to systems of fermions, its authors Ceperley, Chester, and Kalos, simply stated that \textit{`Monte Carlo methods lend themselves easily to more complex problems'}\cite{che+77prb}.

The long list of modern QMC methods can be divided into two main categories. 
The first category includes the methods that are based on the stochastic sampling of the system's degrees of freedom, such as the particles' coordinates, and includes the most famous variational Monte Carlo (VMC)\cite{mcm+65pr,cep+77prb}, diffusion Monte Carlo (DMC)\cite{fou+01rmp} and Lattice-Regularized diffusion Monte Carlo (LRMDC)\cite{cas+10jcp} methods.
The second category, on the other hand, contains all the QMC methods based on the sampling of the configurational space, such as Auxiliary-Field QMC\cite{zha+03prl}, Full Configuration Interaction QMC (FCIQMC)\cite{boo+09jcp,cle+10jcp}, and many others\cite{fou+01rmp,aus+12cr}. \\
Within this vast ensemble of QMC methods\cite{dellapia+25jcp} many excellent open and restricted access QMC codes have been developed, such as TurboRVB\cite{TurboRVB}, QMCPACK\cite{kent2020,Kim2018}, CASINO\cite{Needs2020}, QWALK\cite{Wagner09}, CHAMP\cite{Champ}, PyQMC\cite{PyQMC}, QMC=Chem\cite{QMCChem} and HANDE-QMC\cite{Spencer19}.

In this work, we present the first version of the {\sffamily QMeCha} (Quantum MeCha /'m\epsi k\schwa /) \footnote{The term \textit{mecha} /'m\epsi k\schwa / is the abbreviation of \textit{mechanical}, which in Japanese manga and anime culture refers to all the piloted mechanical devices (robots, cars, airships, computers, and others) introduced in the science fiction genre.}
QMC package, initiated in 2017 as a flexible tool to study many-particle Hamiltonians with different wavefunctions' approximations.
In the years that followed, the code progressively grew to describe interactions between fermions of different charge and mass, such as positrons, muons and anti-muons, interacting with molecules\cite{cha+22jctc,cha+22cs}, and to include a QMC embedding procedure in which a molecular system of nuclei and fermions can be immersed in an environment approximated by point charges and quantum Drude oscillators (QDOs)\cite{Ditte2023,Ditte2024,Ditte2025}. 
The full Hamiltonian that comprises the two types of quantum particles, \textit{ie} fermions and drudons, is integrated over one correlated wavefunction that is constructed as the product of various fermionic terms, such as the Slater determinant, and Jastrow factors. 
Within the code, the observables are computed through an efficient implementation of the VMC method, and through an improved size-consistent DMC algorithm, that reduces finite time-step errors when computing energy differences\cite{dellapia+25jcp,and+21jcp,and+24jcp}.

In this review, the main features of {\sffamily QMeCha} are extensively presented through several examples.
The review is organized in the following way: In Section \ref{sec:mpsyss} we describe the Hamiltonian that the code is able to integrate, in Section \ref{sec:trialwvf} we describe the variational wavefunctions that are used to approximate the Hamiltonian's ground state, in Section \ref{sec:qmc_methods} we discuss the basic features of our QMC methods' implemented, and in Section \ref{sec:comp_eff} we discuss computational details and the efficiency of the basic calculations in {\sffamily QMeCha}.

Finally, {\sffamily QMeCha} has been used to tackle three main sets of applications reported in Section \ref{sec:appl}. 
The first application focuses on the description of van der Waals (vdW) interactions in macromolecules\cite{Puleva2025}, where the DMC calculations are used as state-of-the-art references together with the LNO-CCSD(T)\cite{nagy+17jcp,LocalCC3,LocalCC4,MRCC} method, to construct a dataset for the description of pocket-ligand interactions in proteins. 
The second application is related to positronic chemistry, an emerging field dedicated to the investigation of the still vastly unexplored bound metastable  states of positrons with atomic and molecular systems, before electron-positron pair annihilation\cite{Natisin2017,Mitroy2002,Gribakin2010}. 
It has been shown that QMC methods are a promising tool for studying electron-positron bound states, because they can explicitly describe the electrostatic correlation effects between electron–positron pairs through wavefunctions including the two particle distances\cite{schrader_RAiCC_2_163_1997,bre+98jcp,Kita2010,cha+22jctc,simula_PRL_129_166403_2022,upa+24jctc, cassella_NC_15_5214_2024}, unlike many quantum-chemical post-HF methods (like MP, CC, and CI) that have to rely on extremely large single-particle basis sets. 
The third set of applications is that of the El-QDO embedding method\cite{Ditte2023,Ditte2024,Ditte2025} first introduced in ref. \citenum{Ditte2023} to efficiently capture the quantum correlation effects between an environment modelled through QDOs and point charges\cite{Martyna2006,Martyna2013,Martyna2019} and a molecular subsystem of interest. 
In this application we have focused on the evaluation of solvation energies, excitation energies and bond energy variations of a set of dimers in water clusters.\cite{Ditte2023,Ditte2025} 

In the last Section \ref{sec:concl}, the review is concluded with final remarks and discussions about open challenges and future implementations. 

\begin{figure*}[t!]
\centering
\includegraphics[width=0.9\textwidth]{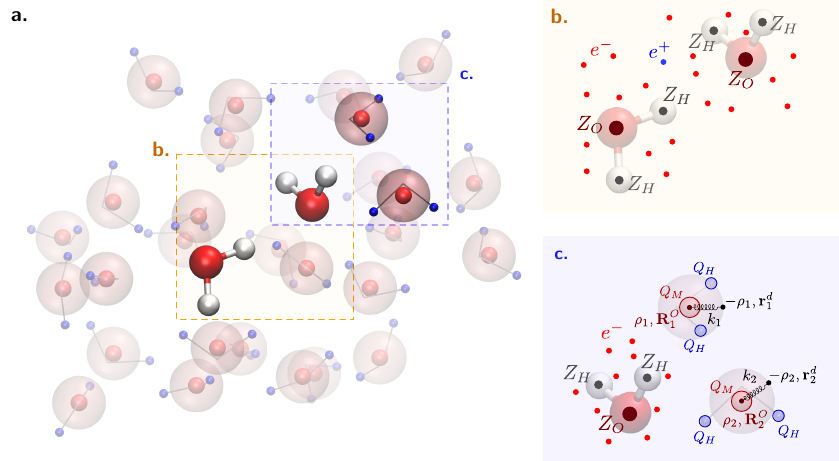}
\caption{
\textbf{a.} Schematic representation of a cluster of 30 water molecules\cite{rak+19jcp} from \href{https://sites.uw.edu/wdbase/database-of-water-clusters/}{sites.uw.edu/wdbase} in which the central water dimer is represented through the fermionic Hamiltonian (\textbf{b}) while the rest of the molecules are substituted by a model obtained from point charges ($Q_{H}$ and $Q_M$) and quantum Drude Oscillators (QDOs) (\textbf{c}).
In \textbf{b} also a positron is displayed to indicate that the code can handle also different types of fermions.
In \textbf{c} the quadratic potential between the QDO center and the drudon is represented as a spring. All the other interactions are described by Coulomb potentials.}
\label{fig:fig1}
\end{figure*}

\section{Many-particle Hamiltonian}\label{sec:mpsyss}

The {\sffamily QMeCha} package has been designed to integrate many-particle Hamiltonians that include fermions\cite{cha+22jctc}, with variable charge and mass, and distinguishable particles such as drudons.
In the next sections, we will briefly describe the single components of the total Hamiltonian that the code can describe, that is represented in Figure \ref{fig:fig1}, explaining their characteristic degrees of freedom.

\subsection{Fermionic Hamiltonian}\label{ssec:fh}

{\sffamily QMeCha} was first implemented to study non-periodic molecular systems.
The first advancement was the explicit inclusion of positronic particles in order to describe the meta-stable states that positrons can form with usually polar molecules.\cite{Natisin2017,Mitroy2002,Gribakin2010,cha+22jctc,cha+22cs}
These systems of molecules and positrons are described considering a general set of $N_f$ fermions with charges and masses $\{q_i,m_i\}_{i=1}^{N_f}$ and coordinate vectors $\bar{\textbf{r}}^f=\{\textbf{r}_i\}_{i=1}^{N_f}$, that include the three dimensional Cartesian coordinates, and a set of $N_n$ nuclei with charges and masses $\{Z_a,M_a\}_{a=1}^{N_n}$ and coordinate vectors $\bar{\textbf{R}}=\{\textbf{R}_a\}_{a=1}^{N_n}$.

Since this version of {\sffamily QMeCha} considers the nuclei as fixed point charges, the masses $\{M_a\}_{a=1}^{N_n}$ are not explicitly considered, and the vector $\bar{\textbf{R}}$ is assumed to be a set of parameters of the system. 
Thus, within the Born-Oppenheimer approximation\cite{bor+27ap} the fermionic Hamiltonian (in atomic units, $\hbar=1$, $4\pi \epsilon_0=1$, $m_e = 1$, $q_e = -1$) is written as
\begin{equation}
\hat{\text{H}}_f=-\sum_{i=1}^{N_f}\frac{1}{2m_i}\nabla_{\textbf{r}_i}^2+\text{V}(\bar{\textbf{r}};\bar{\textbf{R}}),
\label{equ:ferm_ham}
\end{equation}
the sum of the kinetic energy operator of the fermions plus the Coulomb potential $\text{V}(\bar{\textbf{r}},\bar{\textbf{R}})$ containing the interactions between all the charged particles in the system:
\begin{multline}
\text{V}(\bar{\textbf{r}};\bar{\textbf{R}})
=\sum_{a=1}^{N_n}\sum_{i=1}^{N_f}\frac{Z_a q_i}{\left|\textbf{r}_i-\textbf{R}_a \right|} +  \\ 
+\sum_{j>i=1}^{N_f} \frac{q_i q_j}{\left|\textbf{r}_i-\textbf{r}_j \right|}
+\sum_{b>a=1}^{N_n} \frac{Z_a Z_b}{\left|\textbf{R}_a-\textbf{R}_b \right|}.
\label{equ:coulomb}
\end{multline}
We must point out that, within this public release of {\sffamily QMeCha} the mass of the fermionic particles is always assumed to be unitary $m_i=1$, and the fermionic charges are $q_i=\pm1$.
The nuclear charges $\{Z_a\}_{a=1}^{N_n}$ are also integer numbers in atomic units.

\subsection{The Hamiltonian of Quantum Drude Oscillators}\label{ssec:qdo}

The Quantum Drude Oscillator (QDO) model for long-range interactions is obtained from the fluctuation-dissipation theorem\cite{Kubo1966}, which expresses the correlation energy of a system in terms of the charge response function and is at the foundations of the many-body dispersion (MBD) method\cite{Tkatchenko2012,Tkatchenko2017rev}.
Within MBD, Coulomb interactions between oscillators are approximated by first-order dipole-dipole interactions\cite{Tkatchenko2012,Tkatchenko2017rev} and, through correlation effects that are described in terms of the eigenstates of Drude oscillators, the model is able to capture the binding energy components that arise in van der Waals interactions with a lower number of degrees of freedom \cite{Tkatchenko2012,Tkatchenko2017rev}.

Coulomb interacting QDOs have previously been applied within the framework of DMC and path integral Monte Carlo (PIMC) to study the dispersion interactions in noble gas dimers\cite{Martyna2013}, solid\cite{Martyna2009} and fluid xenon\cite{Martyna2006}. 
Furthermore, together with external point charges, a QDO-based model of the water molecule \cite{MartynaPRL2013} has been applied to study the dynamics of liquid water\cite{Martyna2019, MartynaMolPhys2013}. 

Within the Born-Oppenheimer approximation\cite{bor+27ap}, each QDO consists of two particles, a \text{center} of charge $\rho>0$ and a quantum particle with opposite charge and mass $\mu$, the drudon, that interact via a quadratic potential $v\left(r^{Od}\right)=\frac{1}{2}\mu\omega^2 \left( r^{Od} \right)^2$,  
where the frequency $\omega$ determines the slope of the quadratic well and $r^{Od}=|\textbf{R}^{O}-\textbf{r}^d|$ is the Euclidean distance between the drudon's coordinates' vector $\textbf{r}^d$ and its center's $\textbf{R}^{O}$. 
This set of parameters $\{\mathbf{R}_{i}^{O},\rho_{i},\mu_{i},\omega_{i} \}_{i=1}^{N_{O}}$ that characterize each QDO are chosen in such a way to reproduce polarizabilities and dispersion coefficients of atoms or molecules \cite{Martyna2013}.

For a system of $N_O$ interacting QDOs (and thus $N_O$ drudons), the Hamiltonian is written as the sum
\begin{equation}
\hat{\text{H}}_{O}=\sum_{i=1}^{N_{O}}\hat{h}_{i}^{O} 
+
\sum_{j >i=1}^{N_{O}} \left ( \frac{\rho_{i}\rho_{j}}{\left|\mathbf{r}_{i}^{d}-\mathbf{r}_{j}^{d} \right|}
+\frac{\rho_{i}\rho_{j}}{\left|\mathbf{R}_{i}^{O}-\mathbf{R}_{j}^{O} \right|} \right),
\label{eq:H^O}
\end{equation} 
of a single drudon operator $\hat{h}_{i}^{O}$ and the Coulomb interaction potentials between all drudonic pairs and between the pairs of QDO centers.

Each one-body operator is written as the sum 
\begin{equation}
\hat{h}_{i}^{O} = -\frac{1}{2\mu_{i}}\nabla^{2}_{\mathbf{r}_{i}^{d}} + v_i\left(r^{Od}_{i}\right) - \sum_{j (\neq i)=1}^{N_{O}}\frac{\rho_{i}\rho_{j}}{\left|\mathbf{r}_{i}^{d}-\mathbf{R}_{j}^{O} \right|},    
\end{equation}
of the kinetic energy of the drudon, the quadratic potential describing the interaction with its center, and the Coulomb potential describing its interaction with the centers of all the other QDOs.

\subsection{Fermionic coupling to Quantum Drude Oscillators}\label{ssec:fcqdo}

An electronic system containing $N_{n}$ nuclei and $N_{f}$ fermions can be embedded into a bath of $N_{O}$ QDOs giving us the total Hamiltonian of the form
\begin{equation}
\hat{\text{H}}_{tot}=\hat{\text{H}}_{O}+\hat{\text{H}}_{f}+\text{V}_{O,f}, 
\label{eq:H^tot}
\end{equation}
where $\hat{\text{H}}_{O}$ is QDO Hamiltonian defined in eq. \ref{eq:H^O}, $\hat{\text{H}}_{f}$ is the standard fermionic Hamiltonian defined in eq. \ref{equ:ferm_ham}
and $\text{V}_{O,f}$ contains the Coulomb interactions between the particles of the embedding potential and those of the electronic system
\begin{multline}
\text{V}_{O,f}
=
\sum_{i=1}^{N_{f}}\sum_{j=1}^{N_{O}}\left(
\frac{q_i\rho_j}{\left|\mathbf{r}_{i}-\mathbf{R}_{j}^{O}\right|} 
- 
\frac{q_i\rho_{j}}{\left|\mathbf{r}_{i}-\mathbf{r}_{j}^{d}\right|} 
\right) + \\ 
+ \sum_{a=1}^{N_{n}}\sum_{j=1}^{N_{O}}\left(\frac{\rho_{j}Z_{a}}{\left|\mathbf{R}_{a}-\mathbf{R}_{j}^{O}\right|}- \frac{\rho_{j}Z_{a}}{\left|\mathbf{R}_{a}-\mathbf{r}_{j}^{d}\right|} \right).
\label{eq:v_Of}
\end{multline}

\subsection{External point charges as classical embedding}\label{ssec:ff}

To simulate the effects on an atomic and positronic subsystem of an environment of molecules with an intrinsic dipole, such as the water clusters in Fig. \ref{fig:fig1}, in the El-QDO embedding method\cite{Ditte2023,Ditte2025} together with the QDOs, it is possible to introduce fractional point charges\cite{Martyna2019} that act as a classical static Force-Field with a similar formalism to the one used in classical Force-Fields such as GAFF\cite{GAFF} and CHARMM\cite{CGFF}.

Molecular Dynamics (MD) is not implemented in the code yet, and this will be one of the possible future developments. 
For now, a given set of point charges can be assigned to a specific molecule to describe intrinsic dipoles, like in the QDO model of water by Martyna and coworkers\cite{Martyna2019} first used in embedding procedures in ref. \citenum{Ditte2023,Ditte2025}. 

Assuming that each pseudo-molecule in the embedding environment has associated only one QDO and many point charges, and that $n^{(i)}_{C}$ are the number of point charges associated to the $i$th pseudo-molecule corresponding to the $i$th QDO, then the full set of interactions within the embedding potential is written as:
\begin{multline}
\text{V}_{O,C}
=
\sum_{j>i=1}^{N_{O}} \left [ \sum_{a=1}^{n^{(i)}_{C}} \sum_{b=1}^{n^{(j)}_{C}} 
\frac{Q_{a} Q_{b}}{\left|\textbf{R}^C_{a}-\textbf{R}_{b}^{C}\right|} + \right . \\ \left . +
\sum_{a=1}^{n^{(i)}_{C}} \left ( 
\frac{Q_{a} \rho_{j}}{\left|\textbf{R}^O_{j}-\textbf{R}_{a}^{C}\right|} -
\frac{Q_{a} \rho_{j}}{\left|\textbf{r}^d_{j}-\textbf{R}_{a}^{C}\right|} 
\right )
\right ] ,
\label{eq:v_OC}
\end{multline}
where the first terms represent the interactions between point charges of different pseudo-molecules, while the second and third terms represent the interactions between the point charges of one pseudo-molecule with the QDO charges of the other ones in the embedding environment.

The interaction potential between the point charges and the systems of fermions and nuclei is written as 
\begin{equation}
\text{V}_{C,f}
=
\sum_{a=1}^{N_{n}}\sum_{j=1}^{N_{C}}\frac{Z_{a} Q_{j}}{\left|\textbf{R}_{a}-\textbf{R}_{j}^{C}\right|}
+\sum_{i=1}^{N_{f}}\sum_{j=1}^{N_{C}}\frac{q_iQ_{j}}{\left|\textbf{r}_{i}-\textbf{R}_{j}^{C}\right|} .
\label{eq:v_Cf}
\end{equation}

\subsection{Screening of the interactions}\label{ssec:int_scree}

Both interaction potentials in eqs. \ref{eq:v_Of}, \ref{eq:v_OC}, and \ref{eq:v_Cf} present divergences when a fermionic particle approaches the particles in the QDOs (centers and drudons) or the point charges present in the embedding potential, or when the drudons approach the molecular nuclei, and the point charges on the other QDOs.
In order to cure such divergences the total wavefunction of the system must include the description of all the cusps' conditions as described in Section \ref{sssec:cusps}.
Yet, the correct description of the cusps, is not able to properly account for the correct behaviour of the short range interactions between the various particles.
Thus, an alternative, and more efficient approach, that we have used for now only to screen the short-range interaction between the fermions, the two QDO particles and the point charges is that of introducing a damping function to completely remove the divergence of the Coulomb interactions. 
In ref. \citenum{Ditte2025} the study of different screening functions and the effect on the short-range interaction regime, has been presented.
From this study, it was evident that the error function damping, used in previous works\cite{Martyna2019,Ditte2023,Ditte2025}, 
\begin{equation}
\label{Eq:damp_erf}
    f\left(r_{ij},\sigma_{ij}\right)=  \frac{ q_i q_j}{r_{ij}}\text{erf}\left( \frac{r_{ij}} {\sqrt{2}\sigma_{ij}}\right),
\end{equation}
where $q_i$,$q_j$, represent two charges of different particles (nuclei, drudons, fermions) and  $\sigma_{ij}=\sqrt{\sigma_i^2+\sigma_j^2}$ is the cut-off parameter determined by the single cut-offs $\sigma_i$ and $\sigma_j$ of the interacting charges, was the most efficient to guarantee stability and convergence.

\subsection{Atomic pseudopotentials}\label{ssec:apse}

{\sffamily QMeCha} supports also the use of pseudopotentials to substitute the core electrons of atoms, reducing the number of degrees of freedom, and the electronic energy scaling involved in the calculations.
The effective core potentials (ECPs) used, are those written in the standard semi-local form\cite{tra+05jcp1,*tra+14jctc,*tra+15jcp,*kro+16prb,*tra+17jcp,*tra+05jcp2,*ben+17jcp,*ben+18jcp,*ann+18jcp,*bur+07jcp,*bur+08jcp}
\begin{equation}
\hat{\text{V}}^a_{\text{ECP}}(\textbf{r}) 
= 
\text{V}^a_{loc}(|\textbf{r}-\textbf{R}_a|)
+
\sum_{l=0}^{l_{\text{max}}}
\text{V}^a_{l}(|\textbf{r}-\textbf{R}_a|) \hat{\text{P}}^a_l
\end{equation}
where $\hat{\text{P}}^a_l=\sum_{m=-1}^{+l} \left | \text{Y}_{l,m}\rangle \langle \text{Y}_{l,m}\right |$ is the projection operator on the real spherical harmonics centered on the $a$th atom in $\textbf{R}_a$, $\text{V}^a_{loc}(|\textbf{r}-\textbf{R}_a|)$ and $\text{V}^a_{l}(|\textbf{r}-\textbf{R}_a|)$ are local and non-local potentials expanded over a Gaussian set, with $|\textbf{r}-\textbf{R}_a|$ being the distance between the electron and the atomic center. 

If drudons or positrons are present in the system, the use of pseudopotentials is possible in {\sffamily QMeCha}, yet when describing the interactions of these particles with the ECP, the non-local parts are removed, so that for positrons the interaction energy will be equal to 
\begin{equation}
\hat{\text{V}}^a_{\text{ECP}}(\textbf{r}) 
= 
-q_i\text{V}^a_{loc}(|\textbf{r}-\textbf{R}_a|)
\end{equation}
where the negative sign accounts for the positive charge of the positrons, while for the drudons we will have  
\begin{equation}
\hat{\text{V}}^a_{\text{ECP}}(\textbf{r}) 
= 
\rho\text{V}^a_{loc}(|\textbf{r}-\textbf{R}_a|)
\end{equation}
where $\rho$ is the negative charge of the drudon.

The use of only the local part of the ECP to approximate the bare nuclear potential with that of an effective nuclear charge has been seen to introduce only a small bias when computing the interaction energies of positrons and drudons with the pseudoatoms. 
We attribute this to the fact that positrons are repelled by the nuclei and thus do not form a bound state with them. Similarly, the drudons, used to represent the polarization of the embedding potential, rarely come close to the nuclei and remain largely localized in the QDO regions. 

\subsection{External electric polarization field}\label{ssec:polfields}

To study the response properties of systems of electrons, positrons, and drudons, {\sffamily QMeCha} also accepts the addition of an external electric field vector $\textbf{E}$, adding a potential energy contribution to the Hamiltonian, defined as
\begin{equation}
\text{V}_{\textbf{E}}=-\boldsymbol{\mu} \cdot \textbf{E} 
\label{equ:ext_field}
\end{equation}
where $\boldsymbol{\mu}$ is the dipole vector of the system\cite{Stone2013}.

In the most general case of a system of atoms, positrons, together with an embedding system of QDOs and point charges, the total dipole vector can be expressed as the sum of the dipoles of the system of fermions and nuclei 
\begin{equation}
\boldsymbol{\mu}_{f} =\sum_{a=1}^{N_a} Z_a ( \textbf{R}_a -\textbf{R}_c )+\sum_{i=1}^{N_f} q_i(\textbf{r}_i-\textbf{R}_c), 
\label{eq:dip_fn}
\end{equation}
and of the embedding system 
\begin{equation}
\boldsymbol{\mu}_{O} = \sum_{i=1}^{N_O} \rho_i \left ( \textbf{R}^{O}_i - \textbf{r}^{d}_i \right ) +\sum_{i=1}^{N_C} Q_i \textbf{R}^{C}_i.
\end{equation}
In {\sffamily QMeCha} the center on which the dipole is computed, $\textbf{R}_c$, is always the center of the total fixed charges defined as
\begin{equation}
\textbf{R}_c = \frac{\sum_{a=1}^{N_a} Z_a \textbf{R}_a +\sum_{i=1}^{N_O} \rho_i\textbf{R}^O_i + \sum_{i=1}^{N_C} Q_i\textbf{R}^C_i}{\sum_{a=1}^{N_a} Z_a +\sum_{i=1}^{N_O} \rho_i + \sum_{i=1}^{N_C} Q_i}
\end{equation}
that does not contribute if the total charge is null, \textit{ie.} $Q=\sum_{a=1}^{N_a} Z_a +\sum_{i=1}^{N_f} q_i+\sum_{i=1}^{N_C} Q_i=0$. 
Notice that the QDOs never contribute to the total charge, and usually, also the sum of the point-charges charges, used to describe neutral systems with intrinsic dipoles, are null, \textit{ie} $\sum_{i=1}^{N_C} Q_i=0$. 

\begin{figure*}[t!]
\centering
\includegraphics[width=0.9\textwidth]{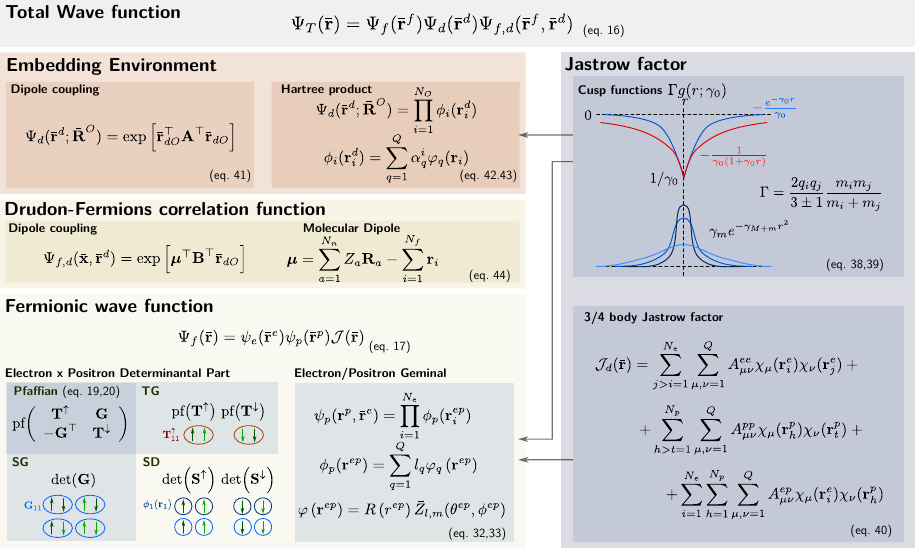}
\caption{
Schematic representation of the total wavefunction presented in the code (eq. \ref{eq:tot_wf}). The Pfaffian includes both Singlet (SG) and Triplet (TG) geminal correlations. Both SG and TG reduce to the Slater determinant (SD) with a particular set of constraints (Section \ref{ssec:eleorposwf}). These wavefunction can be used to describe the fermionic behavior of the separate populations of electrons and positrons (eq. \ref{eq:elecpos_wvf2}).
The cusp functions in the Jastrow factor are used for all sets of particles in the Hamiltonian, nuclei, fermions, QDO centers and drudons.
}
\label{fig:fig2}
\end{figure*}

\section{Trial wavefunctions}\label{sec:trialwvf}

As discussed in the introduction, one of the main advantages of QMC is the possibility to use flexible parametric wavefunctions able to explicitly introduce correlation terms between the quantum particles.

In {\sffamily QMeCha}, in order to describe the quantum state of the many-particle types of the general system, the wavefunction is written as a product of various terms
\begin{equation}
\Psi_T(\bar{\textbf{r}}^f,\bar{\textbf{r}}^d)= \Psi_f(\bar{\textbf{r}}^f) \Psi_d(\bar{\textbf{r}}^d) \Psi_{f,d}(\bar{\textbf{r}}^f,\bar{\textbf{r}}^d), 
\label{eq:tot_wf}
\end{equation}
which are respectively the wavefunction of the fermionic particles (electrons and positrons) $\Psi_f(\bar{\textbf{r}}^f)$, the wavefunction of the purely drudonic particle $\Psi_d(\bar{\textbf{r}}^d)$, when present,  and the coupling function between the fermionic system and the QDO environment, \textit{i.e.} $\Psi_{f,d}(\bar{\textbf{r}}^f,\bar{\textbf{r}}^d)$\cite{Ditte2023}.
A schematic representation of the trial wavefunction present in the code is shown in Fig. \ref{fig:fig2}.

Within the code, no explicit dependency on the spin is present, and we omit to include those degrees of freedom. 
The vector $\bar{\textbf{r}}_f$, thus, represents only the $3N_f$ Cartesian coordinates of the electronic and positronic particles, while the $\bar{\textbf{r}}^d$ represents those of the $3N_d$ drudons.
Therefore, the spin symmetry of the total wavefunction is only taken into account by imposing the correct symmetry of the spatial part of the wavefunction.

In the following sections, we will describe the various components and the different options that can be used to construct each term.

\subsection{Wavefunctions for electrons and positrons}\label{ssec:eleorposwf}

The most general wavefunction $\Psi_f(\bar{\textbf{r}})$ for many electrons and many positrons must explicitly describe the many-body correlation effects between the $3N_e$ electronic Cartesian coordinates $\bar{\textbf{r}}^e$ and the $3N_p$ positronic $\bar{\textbf{r}}^p$ ones in the field of the nuclei $\bar{\textbf{R}}$. 
Yet, this explicitly correlated form is in practice difficult to treat, and various approximations have been introduced in the literature, such as that proposed by Bressanini \textit{et al.} in ref. \citenum{bre+98pra}, which is built as the product (or a linear combination of symmetrized products) of two-particle functions, including all particle pairs' correlations (electron-electron, electron-positron, nucleus-electron, and nucleus-positron) correctly symmetrized (or anti-symmetrized) to take into account the distinguishability (or indistinguishability) of the particles with opposite (or identical) charge and spin.

The simplest possible wavefunction for a system of mixed electrons and positrons is written as the product of two functions (determinants, Pfaffians, or linear combinations of them)  
\begin{equation}
\Psi_f(\bar{\textbf{r}}) = 
\psi_e(\bar{\textbf{r}}^e)
\psi_p(\bar{\textbf{r}}^p) 
\mathcal{J}(\bar{\textbf{r}}) 
\label{eq:elecpos_wvf2}
\end{equation}
each depending only on the electronic $\psi_e(\bar{\textbf{r}}^e)$ or positronic $\psi_p(\bar{\textbf{r}}^p)$ degrees of freedom\cite{kit+09jcp,jor+14pra,cha+18ang,rey+19ijqc,shu+20jcp}.
In this case, the correlation effects between electrons and positrons are only recovered through the Jastrow factor $J(\bar{\textbf{r}})$ (see Section \ref{ssec:jas}) that includes the description of all the nuclear and fermionic cusp conditions, together with various few-body terms usually limited to three- or four-body correlations, that are fundamental in describing the dynamical correlation between particles. 

In the sections that follow we will describe the types of wavefunctions included in {\sffamily QMeCha} that can be used with basis sets of atomic orbitals to describe $\psi_e(\bar{\textbf{r}}^e)$ and $\psi_p(\bar{\textbf{r}}^p)$. 
We will start with the most general one, \textit{ie} the Pfaffian, to the most common Slater determinant, discussing the relationship between them\cite{baj+06prl,baj+08prb,rub+11cpc,gen+19jcp,gen+20jctc}.

\subsubsection{The Pfaffian}\label{sssec:pfaff}

The Pfaffian, introduced by the mathematician Arthur Cayley\cite{cai+1849} and named after Johann Friedrich Pfaff, is defined as the polynomial of the elements of a skew-symmetric matrix with an even number of elements per dimension (for a matrix with odd number of elements per dimension the Pfaffian is zero).  
For a skew-symmetric matrix $\textbf{P}$ with $2n\times 2n$ elements $p_{ij}$, we can define a set $\Pi$ of $(2n-1)!!$ partitions of the $2n$ elements into pairs $\pi = \left [ (i_1,j_1),(i_2,j_2), \ldots, (j_n,j_n)\right]$ with $i_k< j_k$ and $i_1< i_2 < \ldots < i_n$\cite{rub+11cpc}.
The Pfaffian of the matrix $\textbf{P}$ will be given by the sum over the products of the matrix elements organized according to the full set of possible partitions 
\begin{equation}
\text{pf}[\textbf{P}] = \frac{1}{2^n n!} \sum_{\pi \in \Pi} \text{sgn}(\pi) p_{i_1,j_1} p_{i_2,j_2} \ldots p_{i_n,j_n}
\end{equation}
each multiplied by the sign of the corresponding elements' permutation\cite{baj+06prl,baj+08prb,gon+11cpc,rub+11cpc}.
For example in the case of a $4 \times 4$ matrix\cite{baj+06prl,baj+08prb} the Pfaffian corresponds to:
\begin{multline*}
\text{pf} \left ( 
\begin{array}{cccc}
    0   & p_{12}  & p_{13}  & p_{14} \\
-p_{12} &    0    & p_{23}  & p_{24} \\
-p_{13} & -p_{23} &   0     & p_{34} \\
-p_{14} & -p_{24} & -p_{34} & 0      \\
\end{array}
\right ) = \\ = \frac{1}{8} \left ( p_{12}p_{34} - p_{13}p_{24} + p_{14}p_{23} \right ).
\end{multline*}

To describe the Pfaffian wavefunction, let us consider a system with an odd number of electrons $N_e$\cite{baj+06prl,baj+08prb} (the description for the case of $N_p$ positrons is equivalent), for which
\begin{equation}
\psi_e(\bar{\textbf{x}}^e)= \text{pf} [ \textbf{P} ] .
\end{equation}
Let us assume that the electrons are `ordered' with respect to their spin so that the first $N_e^{\uparrow}$ have spin up while the last $N_e^{\downarrow}$ have an opposite spin; by doing so, we can partition the skew-symmetric matrix into four blocks
\begin{equation}
\textbf{P} =
\left (
\begin{array}{cc}
\textbf{T}^{\uparrow} & \textbf{G} \\
-\textbf{G}^{\top} & \textbf{T}^{\downarrow} \\
\end{array}
\right ), 
\label{eq:P}
\end{equation}
where the elements of each block describe the pairing of a subset of electrons in a fixed spin state.
The elements of the $\textbf{T}^{\uparrow}$ and $\textbf{T}^{\downarrow}$ block matrices describe the coupling between electrons with parallel spins in a triplet state, and are defined through the linear combinations of products of two atomic orbitals $\varphi_q(\textbf{x})$ modulated by a set of coupling coefficients $\zeta^{\uparrow}_{qp}$ and $\zeta^{\downarrow}_{qp}$ 
\begin{equation}
\textbf{T}^{\uparrow }_{ij} = \phi_T ( \textbf{r}^{\uparrow}_i, \textbf{r}^{\uparrow}_j ) = \sum_{q,p=1}^{Q} \zeta^{\uparrow }_{qp}\varphi_q(\textbf{r}^{\uparrow}_i)\varphi_p(\textbf{r}^{\uparrow}_j)
\left | 1, 1 \right \rangle
\label{eq:triplet_up}
\end{equation}
\begin{equation}
\textbf{T}^{\downarrow}_{ij} = \phi_T ( \textbf{r}^{\downarrow}_i, \textbf{r}^{\downarrow}_j ) = \sum_{q,p=1}^{Q} \zeta^{\downarrow}_{qp}\varphi_q(\textbf{r}^{\downarrow}_i)\varphi_p(\textbf{r}^{\downarrow}_j)
\left | 1, -1 \right \rangle
\label{eq:triplet_down}
\end{equation}
where the indexes $q$ and $p$ run on the full set of basis set orbitals $Q$, and $\left | 1, 1 \right \rangle$ and $\left | 1, -1 \right \rangle$ correspond respectively to the triplet spin states $\left | \frac{1}{2}, \frac{1}{2}\right \rangle \left | \frac{1}{2}, \frac{1}{2} \right \rangle$ and $\left | \frac{1}{2}, -\frac{1}{2}\right \rangle \left | \frac{1}{2}, -\frac{1}{2} \right \rangle$.
In order to fix the spin symmetry, the spatial part of the geminal functions must be antisymmetric with respect to the exchange of the electronic coordinates, thus the conditions that $\zeta^{\uparrow }_{qp} = -\zeta^{\uparrow }_{pq}$ and $\zeta^{ \downarrow}_{qp} = -\zeta^{ \downarrow}_{pq}$ must always hold. 
Moreover, since two electrons with parallel spin cannot occupy the same orbital, it must be that $\zeta^{\uparrow}_{qq}=0$ and $\zeta^{\downarrow}_{qq}=0$ $\forall\ q \in Q$.
Because of these two conditions, it is clear that the set of coefficients $\zeta^{\uparrow}_{qp}$ and $\zeta^{ \downarrow}_{qp}$ form two skew-symmetric matrices $\mathcal{Z}^{\uparrow}$ and $\mathcal{Z}^{\downarrow}$. 

The elements of the $\textbf{G}_{ij}$ block matrix describe the coupling of electrons with opposite spin in a singlet state $\left | 0,0 \right \rangle = \frac{1}{\sqrt{2}} \left ( \left | \frac{1}{2}, \frac{1}{2}\right \rangle \left | \frac{1}{2}, -\frac{1}{2} \right \rangle - \left | \frac{1}{2}, -\frac{1}{2}\right \rangle \left | \frac{1}{2}, \frac{1}{2} \right \rangle \right )$ through the linear combination of products of two atomic orbitals modulated by the coupling coefficients $\lambda_{qp}$:
\begin{equation}
\textbf{G}_{ij} = \phi_G ( \textbf{r}^{\uparrow}_i, \textbf{r}^{\downarrow}_j ) = \sum_{q,p=1}^{Q} \lambda_{qp}\varphi_q(\textbf{r}^{\uparrow}_i)\varphi_p(\textbf{r}^{\downarrow}_j)
\left | 0,0 \right \rangle ,
\label{eq:g}
\end{equation}
where the coefficients $q$ and $p$, as for the triplet geminals in eqs. \ref{eq:triplet_up} and \ref{eq:triplet_down}, run over the full basis set $Q$. 
In order to fix the spin symmetry, the spatial part of the geminal functions must be strictly symmetric, and thus the coefficients have to satisfy the condition $\lambda_{qp} = \lambda_{pq}$ and can be regrouped in a symmetric matrix $\Lambda$.

In the case in which the number of electrons (or positrons) in the system is odd, the Pfaffian wavefunction can be generalized following the procedure proposed by Sorella and coworkers in refs. \citenum{gen+19jcp,gen+20jctc} or by simply adding a column (and a row) of elements, representing the electrons occupying an unpaired molecular orbital\cite{baj+06prl,baj+08prb}. 
In {\sffamily QMeCha} we follow this last scheme, for which 
\begin{equation}
\textbf{P}=
\left (
\begin{array}{ccc}
\textbf{T}^{\uparrow} & \textbf{G} & \bar{\phi}_\uparrow \\
-\textbf{G}^{\top} & \textbf{T}^{\downarrow} & \bar{\phi}_\downarrow  \\
-\bar{\phi}^{\top}_\uparrow &  -\bar{\phi}^{\top}_\downarrow & 0 \\
\end{array}
\right ),
\label{eq:P_g}
\end{equation}
where $\textbf{T}^{\uparrow }$, $\textbf{T}^{\downarrow }$ and $\textbf{G}$ are the matrices defined above, while the elements of the last column  (or row) vectors
\begin{equation}
\bar{\phi}_i^\uparrow =
\phi^{\uparrow} (\textbf{r}^{\uparrow}_i) = \sum_{q}^{Q} l^{\uparrow}_q \varphi_q(\textbf{r}^{\uparrow}_i) \qquad i \in N^{\uparrow}_e\\
\end{equation}
and
\begin{equation}
\bar{\phi}_i^\downarrow =
\phi^{\downarrow} (\textbf{r}^{\downarrow}_i) = \sum_{q}^{Q} l^{\downarrow}_q \varphi_q(\textbf{r}^{\downarrow}_i) \qquad i \in N^{\downarrow}_e\\
\end{equation}
represent the values of molecular orbitals, occupied by the spin up or spin down electrons (or positrons), and depending on the linear coefficients $l^{\downarrow}_q$ and $l^{\uparrow}_q$ that are in general different for the two spin populations. 

\subsubsection{The antisymmetrized geminal power}\label{sssec:agp}

To recall the relationship between the Pfaffian wavefunction and antisymmetrized geminal power (AGP) wavefunction\cite{col+63rmp,*col+65jmp,cas+03jcp,cas+04jcp} let us consider a closed shell system with an even number of electrons. 
By removing in the $\textbf{P}$ matrix defined in eq. \ref{eq:P}, the coupling between the electrons with parallel spin ($\mathcal{Z}^{\uparrow} = \mathcal{Z}^{\downarrow} = 0$), the Pfaffian reduces to the determinant of the geminal matrix $\textbf{G}$\cite{baj+06prl,baj+08prb},
\begin{equation}
\psi_e(\bar{\textbf{x}}^e) = 
\text{pf} 
\left (
\begin{array}{cc}
0 & \textbf{G} \\
-\textbf{G}^{\top} & 0 \\
\end{array}
\right ) = 
\text{det} 
\left [
\textbf{G} 
\right ],
\end{equation}
which corresponds to the AGP wavefunction\cite{col+63rmp,*col+65jmp,cas+03jcp,cas+04jcp} for closed shell systems.
The elements of the geminal matrix are thus those defined in eq. \ref{eq:g} and describe the coupling between electrons of opposite spin in a pure singlet state.

For a spin polarized systems $(N^{\uparrow}_e > N^{\downarrow}_e)$ the geminal matrix can be generalized\cite{col+65jmp,cas+03jcp} by adding $N^u_e = N^{\uparrow}_e - N^{\downarrow}_e$ columns, each with $N_e^\uparrow$ elements, containing unpaired orbitals
\begin{equation}
\textbf{G}_{ij} = \phi_j( \textbf{r}^{\uparrow}_i) = \sum_{q=1}^{Q} l^{j}_{q} \varphi_{q}  ( \textbf{r}^{\uparrow}_i ) 
\quad
\begin{array}{l}
i \in [1,N^{\uparrow}_e] \\
j \in [N^{\downarrow}_e+1, N^{\uparrow}_e]\\
\end{array},
\end{equation}
occupied solely by the spin-up electrons: In this way we reconstruct a square $\textbf{G}$ matrix of $N^{\uparrow}_e \times N_e^{\uparrow}$ elements.
 
\subsubsection{The Slater determinant}\label{sssec:sd}

The AGP wavefunction is a constrained multideterminantal expansion\cite{mor+12jctc,zen+13jctc,bar+15jctc,bar+15jctc_b} thus containing the single Slater determinant in its variational space. 
The relationship between the AGP and the Slater determinant can be understood by considering that the geminal functions in eq. \ref{eq:g}, with an appropriate transformation\cite{baj+08prb,mar+09jcp}, can be rewritten as 
\begin{equation}
\phi_G ( \textbf{r}^{\uparrow}_i, \textbf{r}^{\downarrow}_j ) = \sum_{k=1}^{Q} \tilde{\lambda}_{k}\phi_k(\textbf{r}^{\uparrow}_i)\phi_k(\textbf{r}^{\downarrow}_j)
\left | 0,0 \right \rangle ,
\end{equation}
where $\phi_k(\textbf{r})=\sum_{q=1}^Q l^{k}_{q}\varphi_q(\textbf{r})$ are a set of $Q$ molecular orbitals (doubly occupied by the electronic pair in a singlet state) and $\tilde{\lambda}_{k}$ are the set of linear coefficients that weight the doubly occupied orbitals in the expansion.
If we restrict the summation to the minimum number of doubly occupied molecular orbitals in the system, \textit{i.e.} $N_e/2$, the geminal reduces to the sum $\phi_G ( \textbf{r}^{\uparrow}_i, \textbf{r}^{\downarrow}_j ) = \sum_{k=1}^{N_e/2} \phi_k(\textbf{r}^{\uparrow}_i)\phi_k(\textbf{r}^{\downarrow}_j) \left | 0,0 \right \rangle$, where all the weights are $\tilde{\lambda}_{k}=1$ $\forall\ k \leq N_e/2$ and $\tilde{\lambda}_{k}=0$ $\forall\ k > N_e/2$.
In this situation, we obtain a reduced geminal matrix that corresponds to the matrix product $\textbf{G}^*=\textbf{S}^{\uparrow \top}\textbf{S}^{\downarrow}$ between the square matrices $\textbf{S}^{\uparrow}$ and $\textbf{S}^{\downarrow}$ containing respectively the values of the $N_e/2$ molecular orbitals computed on the $N_e^{\uparrow}=N_e/2$ spin up and $N_e^{\downarrow}=N_e/2$ spin-down electrons. 

This reduced geminal matrix corresponds to the Slater determinant wavefunction written as the product of the two determinants of $\textbf{S}^{\uparrow}$ and $\textbf{S}^{\downarrow}$
\begin{equation}
\psi_e(\bar{\textbf{x}}^e) = \text{det} \left [ \textbf{S}^{\uparrow \top}\textbf{S}^{\downarrow}\right ] = \text{det} \left [ \textbf{S}^{\uparrow}\right ]\text{det} \left [ \textbf{S}^{\downarrow}\right ] .
\end{equation}

\subsection{Electron-Positron Correlations}\label{ssec:eleposgem}

The wavefunctions listed in subsection \ref{ssec:eleorposwf} can all be used to represent, through atomic basis sets, the wavefunction of many electrons and many positrons separately, as the product in eq. \ref{eq:elecpos_wvf2}.

The limitation of this type of factorization lies in the fact that, while atomic basis sets are a good basis for electrons that form bound states with the nuclei, they are not an appropriate representation for positrons that are repelled by the atomic nuclei and attracted by the electrons with which they can form metastable states before annihilation\cite{Natisin2017,Gribakin2010,Mitroy2002,Cheng2012,Harabati2014,Charry2018,Moncada2020,Goli2019,shu+20jcp,Bressanini2021.one,Bressanini2021.two}.

Thus, for the case of many electrons and one positron, an alternative way to construct the positronic wavefunction is to assume that while the electronic wavefunction $\psi_e(\bar{\textbf{r}}^e)$ describes the spin and angular symmetries of the electrons in the field of the nuclei, $\psi_p(\textbf{r}^p, \bar{\textbf{r}}^e)$ depends only on the distances between the single positron and electrons, and not on the nuclear coordinates; with the constraint of being symmetric for the exchange of any electronic coordinate.

One way to achieve this is by constructing $\psi_p(\textbf{r}^p, \bar{\textbf{r}}^e)$ through the `Positronium' basis set\cite{bre+98pra,bre+98jcp,cha+22jctc}, which explicitly depends on the electron-positron distance.
As a matter of fact, it can be easily shown that the ground state of a system of one electron and one positron, \textit{i.e.} the Positronium (Ps), can be exactly described by an exponential function of the electron-positron distance $r^{ep}=|\textbf{r}^e-\textbf{r}^p|$:
\begin{equation}
\varphi\left(\textbf{r}^{ep}\right) = r^{ep} R\left(r^{ep}\right)Y_{l}^{m}(\theta^{ep},\phi^{ep}),
\label{eq:ep_orbs}
\end{equation}
where $R\left(r^{ep}\right)$ is a radial function normalized with respect to the distance $r^{ep}$ and $Y_{l}^{m}(\theta^{ep},\phi^{ep})$ is a real spherical harmonic (centered on the positron) that is used to introduce an angular momentum.\cite{bre+98pra,cha+22jctc} 

Through this basis, we can construct a positronic wavefunction for many electrons and one positron as the product:
\begin{equation}
\psi_p(\textbf{r}^p, \bar{\textbf{r}}^e) = \prod_{i=1}^{N_e} \phi_p(\textbf{r}^{ep}_{i}) ,
\label{eq:EPO}
\end{equation}
of identical orbitals (so that the function is symmetric with respect to the exchange of the electronic coordinates), each dependent on the electron-positrons distance $\textbf{r}^{ep}_{i}$, thus referred to as electron-positron orbitals (EPO), that are defined as linear combinations 
\begin{equation}
\phi_p(\textbf{r}^{ep}) = \sum_{q=1}^{Q} l_q \varphi_q\left(\textbf{r}^{ep}\right)
\label{eq:posgem}
\end{equation}
of the newly defined Positronium orbitals.

\subsection{Jastrow factors}\label{ssec:jas}

The bosonic Jastrow factor\cite{boy+69prs,dru+04prb} for electrons and positrons that is implemented in {\sffamily QMeCha}, is written as the exponential of a sum of two main factors
\begin{equation}
\mathcal{J}(\bar{\textbf{r}}) 
= 
\text{exp} \{ 
J_c(\bar{\textbf{r}})
+J_{d}(\bar{\textbf{r}})
\}
\label{EQ:Jastrow}
\end{equation}
both depending on the fermionic and nuclear degrees of freedom (that have been omitted in the equation), of which the first contains an explicit description of the wavefunctions' cusp conditions (both for fermion-nucleus and fermion-fermion), while the second is used to describe the few-body correlation effects between the fermions in the field of the nuclei.
This Jastrow is a generalization\cite{cha+22jctc,cha+22cs} of that introduced in refs. \citenum{cas+04jcp,mar+09jcp}.

In the following subsections, we will describe these two terms and explain their general constructions.

\subsubsection{General cusp functions}\label{sssec:cusps}

The Jastrow $J_c(\bar{\textbf{r}})$ term that is used to describe the many particles' cusp conditions is written as the linear combination 
\begin{equation}
    J_c(\bar{\textbf{x}}) = \sum_{i=1}^{N_f} \sum_{a=1}^{N_n} f_{1b} (r_{iq} )  +\sum_{j>i=1}^{N_f}  f_{2b} (r_{ij})
    \label{eq:Jcusp}
\end{equation}
of one-body terms, describing the fermion-nucleus cusp, and the two-body terms, describing the fermion-fermion cusps,
depending respectively on the distance between a fermion and a nucleus, $r_{ia}=\left | \textbf{r}_i - \textbf{R}_a\right |$, and the distance between two fermions $r_{ij}=\left | \textbf{r}_i - \textbf{r}_j\right |$.

For two general charged particles with finite masses, interacting via the Coulomb potential, Kato's cusp conditions\cite{Kato1951,Kato1957} are written in the form 
\begin{equation}
\frac{1}{\left \langle \Psi \right \rangle } \left . \frac{\partial \left \langle \Psi \right \rangle }{\partial r_{ij}}\right |_{r_{ij}=0} = \frac{2q_i q_j }{3\pm 1}\frac{m_{i}m_{j}}{m_{i}+m_{j}} = \Gamma 
\label{eq:KatoCusp}
\end{equation}
where in the factor $ \frac{1}{3\pm 1}$ the $+$ sign holds for indistinguishable particles and the $-$ holds for distinguishable ones.

For the special case in which one particle is a nucleus with infinite mass (fixed point charge) eq. \ref{eq:KatoCusp} reduces to
\begin{equation}
\frac{1}{\left \langle \Psi \right \rangle } \left . \frac{\partial \left \langle \Psi \right \rangle }{\partial r_{ia}}\right |_{r_{ia}=0} = q_i Z_a m_i .
\end{equation}

In order to satisfy these conditions, in the code we write both the one-body and two-body cusps in the form:
\begin{equation}
f(r;\boldsymbol{\gamma}) = \Gamma g(r; \gamma_0) + \sum_{m=1}^{M} \gamma_m e^{-\gamma_{M+m} r^2}
\label{eq:fun_csp}
\end{equation}
where the functions $g(r; \gamma_0)$ at the moment, can have the forms:
\begin{equation}
g(r; \gamma_0) = 
\left \{
\begin{array}{l}
-\frac{e^{-\gamma_0 r}}{\gamma_0}\cite{cas+03jcp} \\
-\frac{1}{\gamma_0(1+\gamma_0 r)}\cite{Pade1892} \\
\end{array}
\right .
\label{eq:fcs}
\end{equation}
In eq. \ref{eq:fun_csp}, the vector of $2M+1$ parameters $\boldsymbol{\gamma}$ is optimized variationally.
In the case of the fermion-nucleus charge, the parameter $\gamma_0$ is modulated to take into consideration the variation of the nuclear charges, and it is multiplied by the factor $2 Z_a^{1/4}$ where $Z_a$ is the atomic charge, \textit{ie.} $\gamma_0=2 Z_a^{1/4}\tilde{\gamma}_0$ where $\tilde{\gamma}_0$ is now the variational parameter that is optimized.
In the code, each atom has its own independent cusp function, and the parameters are eventually connected by symmetry.

\subsubsection{Dynamical Jastrow factor}\label{sssec:dynJas}

As anticipated at the beginning of section \ref{ssec:jas} the dynamical part $J_{d}(\bar{\textbf{r}})$ of the Jastrow factor is a generalization of that introduced in refs. \citenum{cas+04jcp,mar+09jcp}, based on an expansion in $Q$ non-normalized atomic orbitals $\chi_{\nu}(\textbf{r})$ (the atomic index is included in the orbital index $\mu$ or $\nu$ and will always be omitted in the following sections).\cite{cha+22jctc,cha+22cs}

In the most general case, it is thus expanded as the sum of three groups of terms
\begin{multline}
J_{d}(\bar{\textbf{r}}) = 
\sum_{j>i=1}^{N_e} \sum_{\mu,\nu=1}^{Q} A^{ee}_{\mu\nu} \chi_{\mu}(\textbf{r}^e_i) \chi_{\nu}(\textbf{r}^e_j)  + \\
+ \sum_{h>t=1}^{N_p} \sum_{\mu,\nu=1}^{Q} A^{pp}_{\mu\nu} \chi_{\mu}(\textbf{r}^p_h) \chi_{\nu}(\textbf{r}^p_t) + \\
+ \sum_{i=1}^{N_e}\sum_{h=1}^{N_p} \sum_{\mu,\nu=1}^{Q} A^{ep}_{\mu\nu} \chi_{\mu}(\textbf{r}^e_i) \chi_{\nu}(\textbf{r}^p_h).
\label{eq:dyn_Jastrow}
\end{multline} 
where $A^{pp}_{\mu\nu}$, $A^{ep}_{\mu\nu}$ and $A^{pp}_{\mu\nu}$ are a set of coefficients that can be regrouped into square matrices $\textbf{A}^{ee}$, $\textbf{A}^{pp}$, and $\textbf{A}^{ep}$.

To avoid spin contamination, the $\textbf{A}^{ee}$ and $\textbf{A}^{pp}$ matrices must be symmetric, \textit{ie} $A^{ee}_{\mu\nu}=A^{ee}_{\nu\mu}$ and  $A^{pp}_{\mu\nu}=A^{pp}_{\nu\mu}$, in order for the Jastrow to be invariant with respect to the exchange of two electrons or two positrons, no matter their spin.

These set of terms describes two different fermions correlated to one (three-body terms) or two nuclei (four-body terms).
Since the total number of parameters of this term scales proportional to $Q^2$ and $Q$ grows linearly with the number of atoms in the system, in {\sffamily QMeCha} it is possible to reduce this Jastrow to only the full set of three-body terms, \textit{ie.}  two electrons coupled on the full set of orbitals of one atom (with positive overlap), or only the diagonal elements, \textit{ie.} two fermions occupying the same orbital of the same atom. 
In both cases the number of parameters scales only linearly with the number of atoms in the system.

\subsection{Wavefunction of drudons}\label{ssec:qdowf}

The QDO wavefunction $\Psi_{d}(\bar{\textbf{r}}^d; \bar{\textbf{R}}^c)$ is constructed to depend only on the coordinates of the Drudons and parametrically on the centers of QDOs.

In {\sffamily QMeCha} we have two distinct types of functions that can be used\cite{Ditte2024}.
The first follows the exact solution of the QDO model interacting via dipole potential~\cite{Ditte2024,Martyna2013} and is thus written in the exponential form 
\begin{equation}
\label{eq:psi_d}
\Psi_{d}(\bar{\textbf{r}}^d; \bar{\textbf{R}}^O) =\text{exp}\left [ \bar{\textbf{r}}_{dO}^{\top} \textbf{A}^{\top} \bar{\textbf{r}}_{dO} \right ] 
\end{equation}
where $\bar{\textbf{r}}_{dO}=\bar{\textbf{r}}^d - \bar{\textbf{R}}^O$ is the vector of the $3N_O$ components of the distances between each drudon from its center, and $\textbf{A}$ is the square symmetric matrix containing $3N_{O}\left(3N_{O} +1\right)/2$ independent parameters that explicitly correlates the fluctuations of the different QDOs, within each other. 

An alternative approach to the QDO wavefunction, introduced in ref. \citenum{Ditte2024} and is built as the Hartree product of 
\begin{equation}
\Psi_{d}(\bar{\textbf{r}}^d; \bar{\textbf{R}}^O) = \prod_{i=1}^{N_O} \phi_i(\textbf{r}^d_i ) , 
\label{eq:psi_prod_J}
\end{equation}
where
\begin{equation}
\phi_i(\textbf{r}^d_i) = \sum_{q=1}^Q \alpha^i_q \varphi_q(\textbf{r}^d_i)
\label{eq:m_orbs}
\end{equation}
are molecular orbitals constructed on a basis of atomic Gaussian (around the QDO's center) or Slater or Mixed type of orbitals (Section \ref{ssec:basissets}).
In order to introduce correlations between drudons, the code also implements the one-body functions between the drudonic particles and the opposite QDO centers, with which they interact via the Coulomb potential, and the two-body cusp functions between the drudonic pairs. 
These terms, constructed according to what described in Section \ref{sssec:cusps} are also useful to recover dynamical correlation between particles, greatly enahance the convergence of the drudonic ground state\cite{Ditte2024}.

\subsection{Fermions and QDOs correlation function}\label{ssec:fermqdocoupling}

The last part of the total wavefunction is the coupling function $\Psi_{f,d}(\bar{\textbf{x}},\bar{\textbf{r}}^d)$ between fermions and the QDO environment.
Following the dipole approximation\cite{Ditte2023}, this coupling term is written as
\begin{equation}
\Psi_{f,d}(\bar{\textbf{x}},\bar{\textbf{r}}^d) = \text{exp} \left [ \bm{\mu}^{\top} \textbf{B}^{\top} \bar{\textbf{r}}_{dO} \right ], 
\label{eq:psi_ed}
\end{equation}
where $\bar{\textbf{r}}^{dO}$ is the vector of the distances between each drudon and its corresponding center, defined in the previous section, $\bm{\mu}$ is the 3-dimensional vector of the dipole moment of the total fermionic sub-system described in eq. \ref{eq:dip_fn}, and $\textbf{B}$ is a rectangular coupling matrix containing $3 \times 3N_{O}$ free parameters. 

In the limit in which the fermionic sub-system and the drudonic environment only interact through non-covalent bonds, the term described in eq.~\ref{eq:psi_ed} has the purpose of recovering the dynamical correlation between the electrons and drudons. 
This factor is chosen to be always positive, since it recovers correlation that is responsible for dispersion, polarization, and electrostatic effects that change the nodal structure of the fermionic sub-system only indirectly, through the fermionic wavefunction described in Sections \ref{ssec:eleorposwf}, \ref{ssec:eleposgem} and  \ref{ssec:jas}.

\subsection{Basis sets}\label{ssec:basissets}

The atomic and Positronium orbital basis sets used in {\sffamily QMeCha} are built in the same manner, as the product of a radial part, that only depends on the distance between the electron and the nucleus $\textbf{r}_{a}$, on which the basis set is centred, and a cubic (or tesseral) harmonics:
\begin{multline}
\varphi (\textbf{r}_{a}) =r_a^l R(r_{a}) Z_{l,m}(x_{a}, y_{a}, z_{a} ) = \\= R(r_{a}) \bar{Z}_{l,m}(x_{a}, y_{a}, z_{a} ) ,
\end{multline}
where $\bar{Z}_{l,m}(x_{a}, y_{a}, z_{a} ) = r_a^l Z_{l,m}(x_{a}, y_{a}, z_{a} )$ are `reduced' cubic harmonics.

For a general orbital written as
\begin{equation}
\varphi (\textbf{r}_{a}) = r_a^l R(r_{a}) Z_{l,m}(x_{a}, y_{a}, z_{a} ),
\end{equation}
the radial part of the wavefunction is written as a linear combination of $p$ different primitive functions.
\begin{equation}
R(r_{a})=\sum_{k=0}^p c_k P_{p}(\zeta_k, r_{a}),
\end{equation}
written as
\begin{equation}
P_{n}(r_{a}) = \mathcal{N}_{n,l}(\zeta) r_a^{n} e^{\alpha(\zeta, r_a)},
\end{equation}
where $\mathcal{N}_{n,l}(\zeta)$ is the normalization factor that depends on the type of exponential function $\alpha(\zeta, r_a)$.

In {\sffamily QMeCha} at the moment we have generalized three types of exponential functions that are
\begin{equation}
\alpha(\zeta, r_a) =
\left \{
\begin{array}{ll}
-\zeta r   & \text{Slater type} \\
-\zeta r^2 & \text{Gaussian type} \\ 
-\frac{(\zeta r_a)^2}{1+\zeta r_a} & \text{Mixed type}\cite{pet+10jcp,pet+11jcp} \\ 
\end{array}
\right . .
\end{equation}
The code also contains an additional set of primitive functions without cusp, of the form:
\begin{equation}
P_{n}(r_{a}) = \mathcal{N}_{n,l}(\zeta) r_a^{n} (1+\zeta r_a) e^{-\zeta r_a}.
\end{equation}

For atoms, the orbitals are centered on the nucleus, while in the case of the Positronium basis sets they are centered on the positron.
The orbitals used for the Jastrow factor do not include the normalization of the radial part so that $\mathcal{N}_{n,l}(\zeta)=1$.

\section{Quantum Monte Carlo methods}\label{sec:qmc_methods}

The quantum Monte Carlo (QMC) methods\cite{fou+01rmp,bec+17} implemented in {\sffamily QMeCha} are variational Monte Carlo (VMC) and diffusion Monte Carlo (DMC). 
Within the VMC framework, it is possible to do a full optimization of the variational space using the Stochastic Reconfiguration (SR) approach introduced by S. Sorella in ref. \citenum{sor+00prb,bec+17}. 

The DMC algorithm implemented in {\sffamily QMeCha} is a modification of the efficient algorithm first introduced by S. Umrigar \textit{et al.}\cite{umr+93jcp}, with a size-consistent cut-off and generalized for particles of different flavors.\cite{and+24jcp,and+21jcp,zen+16prb,zen+19jcp} 

In the next sections, we will briefly describe the QMC methods implemented and give some examples of their application. 

\subsection{Variational Monte Carlo}\label{ssec:vmc}  

The VMC method\cite{fou+01rmp,bec+17} consists in the application of the Monte Carlo stochastic integration to the evaluation of the energy functional 
\begin{equation}
\text{E}\left[\Psi_T(\boldsymbol{\alpha})\right]=
\frac{\int \Psi^{*}_T(\bar{\textbf{r}}; \boldsymbol{\alpha}) \hat{\text{H}}\Psi_T(\bar{\textbf{r}}; \boldsymbol{\alpha})d\bar{\textbf{r}} }{\int \left |\Psi_T(\bar{\textbf{r}}; \boldsymbol{\alpha}) \right |^2 d\bar{\textbf{r}} }
\label{equ:E_fun}
\end{equation}
over a variational \textit{trial} wavefunction $\Psi_T(\bar{\textbf{r}}; \boldsymbol{\alpha})$, that approximates an eigenstate, usually the ground state, of the system's Hamiltonian $\hat{\text{H}}$.
For the sake of simplicty, in eq. \ref{equ:E_fun} we have omitted the explicit reference to the spin coordinates since, the Hamiltonians used in {\sffamily QMeCha} do not depend explicitly on the spin, and the spin state is only imposed through the symmetry of the trial wavefunction $\Psi_T(\bar{\textbf{r}}; \boldsymbol{\alpha})$.

In order to perform a Monte Carlo integration of eq. \ref{equ:E_fun}, the integrand is rewritten as the product of two functions of the coordinate vector $\bar{\textbf{r}}$
\begin{displaymath}
\text{E}\left[\Psi_T\right] = \int E_{l}(\bar{\textbf{r}}; \boldsymbol{\alpha}) \Pi(\bar{\textbf{r}}; \boldsymbol{\alpha}) d\bar{\textbf{r}}
\label{equ:E_VMC}
\end{displaymath}
that are respectively the \textit{local energy}
\begin{equation}
\text{E}_{l}(\bar{\textbf{r}}; \boldsymbol{\alpha})=\frac{\hat{\text{H}} \Psi_T(\bar{\textbf{r}}; \boldsymbol{\alpha})}{\Psi_T(\bar{\textbf{r}}; \boldsymbol{\alpha})} ,
\end{equation}
defined as the energy associated to a given configuration $\bar{\textbf{r}}$, and the probability density
\begin{equation}
\Pi(\bar{\textbf{r}}; \boldsymbol{\alpha})=\frac{|\Psi_T(\bar{\textbf{r}}; \boldsymbol{\alpha})|^2}{\int |\Psi_T(\bar{\textbf{r}}; \boldsymbol{\alpha})|^2 d\bar{\textbf{r}}}
\end{equation}
to find the particles in that particular configuration $\bar{\textbf{r}}$.

The stochastic integration is then obtained by sampling a chosen number $\mathcal{N}$ of `uncorrelated' configurations $\bar{\textbf{r}}$ extracted according to the probability density $\Pi(\bar{\textbf{r}}; \boldsymbol{\alpha})$ through the Metropolis-Hastings algorithm\cite{met+53jcp,has+70b}.
For each value of $\bar{\textbf{r}}$, the local functions, such as the local energy $\text{E}_{l}(\bar{\textbf{r}}; \boldsymbol{\alpha})$, are computed and accumulated so that their estimation, and in particular the estimation of the energy functional $\text{E}\left[\Psi_T(\boldsymbol{\alpha})\right]$ reduces to the statistical average
\begin{equation}
\text{E}\left[\Psi_T(\boldsymbol{\alpha})\right]\approx  \bar{\text{E}}_{l}(\boldsymbol{\alpha}) = \left\langle \text{E}_{l}(\boldsymbol{\alpha})\right\rangle_\mathcal{N} = \frac{1}{\mathcal{N}} \sum_{i=1}^{\mathcal{N} }\text{E}_{l}(\bar{\textbf{r}}_i; \boldsymbol{\alpha})
\end{equation}
with the associated statistical error
\begin{equation}
\sigma_{\bar{\text{E}}_l(\boldsymbol{\alpha})} = \sqrt{ \frac{s^2_{\bar{\text{E}}_l(\boldsymbol{\alpha})}}{\mathcal{N}} }, 
\end{equation}
that is proportional to the sample variance of the local energies 
\begin{equation}
s^2_{\bar{\text{E}}_l(\boldsymbol{\alpha})} = \left\langle \text{E}^2_{l}(\boldsymbol{\alpha})\right\rangle_\mathcal{N}-\left\langle \text{E}_{l}( \boldsymbol{\alpha})\right\rangle^2_\mathcal{N}
\label{equ:var_E_loc}
\end{equation}
and decreases as the square root of $\mathcal{N}$.

Here we have indicated with $\langle \cdots \rangle_\mathcal{N}$ as the sample average of the function in the brackets, and from now on we will indicate with $\langle \cdots \rangle_{\Pi(\boldsymbol{\alpha})}$  the population average on the entire distribution of configurations $\bar{\textbf{r}}$.

It is important to notice that, from the definition of the variance in eq. \ref{equ:var_E_loc}, the VMC method presented here satisfies the \textit{zero variance principle}\cite{bec+17}, \textit{ie.} if the trial wavefunction $\Psi_T(\bar{\textbf{r}}; \boldsymbol{\alpha})$ is an eigenfunction of the Hamiltonian\cite{umr+88prl,umr+05prl}, the local energy function $\text{E}_{l}(\bar{\textbf{r}}; \boldsymbol{\alpha})$ becomes a constant that corresponds to the eigenvalue associated to $\Psi_T(\bar{\textbf{r}}; \boldsymbol{\alpha})$, and thus the variance is null.

Both the \textit{variational principle}, for which the energy functional defined in eq. \ref{equ:E_fun} is always an upper bound to the ground state energy of the Hamiltonian, \textit{ie.} $\text{E}\left[\Psi_T(\boldsymbol{\alpha})\right] \ge E_0$, and the zero variance principle have been exploited to optimize the parameters $\boldsymbol{\alpha}$ of the wavefunction\cite{sor+01prb,umr+07prl,tou+07jcp,sor+07jcp}, yet, before discussing the optimization procedure implemented in the code, we want to discuss here some details regarding the VMC sampling.

The extension of the VMC algorithm to particles of different flavours is rather straightforward.  
In our approach, the sets of particles are diffused particle-by-particle in random order starting from the fermions (electrons and positrons), according to the Metropolis-Hastings algorithm\cite{met+53jcp,has+70b}.
Each particle's trial move is proposed according to the transition probability
\begin{equation}
\textbf{r}'_i = \textbf{r}_i + \sqrt{\Delta_i} \boldsymbol{\eta},
\end{equation}
where $\boldsymbol{\eta}$ is a 3-dimensional vector of Gaussian distributed random numbers with zero mean and unitary variance, and $\Delta_i$ is an amplitude that depends on the type of particle and is defined as:
\begin{equation}
\Delta_i= \left \lbrace
\begin{array}{ll}
\delta_e / m_e & \text{for } i \in [1,N_e] \\
\delta_p / m_p & \text{for } i \in [1,N_p] \\
\delta_d / \mu_i & \text{for } i \in [1,N_d] \\
\end{array} \right . .
\end{equation}
The parameters $\delta_e$, $\delta_p$, and $\delta_d$ are amplitudes used respectively for the electrons, positrons, and drudons, while $m_e$, $m_p$ and $\mu_i$ are the corresponding masses. 
Within the code, the $\delta_e$, $\delta_p$, and $\delta_d$ parameters are optimized during the thermalization process by converging the acceptance probability of the MC moves to the value of 50\%, which is the rule of thumb, that has the purpose of balancing the acceptance rate of the single particle moves and the correlation between configurations and thus between local observable evaluations.
In order to further reduce the correlation, the two-step algorithm developed in ref. \citenum{dew+00jcp} is also applied to split the acceptance probability between the determinant part of the total wavefunction plus the one-body Jastrow, and the two- the three- and four body Jastrow factors implemented in the code\cite{cas+03jcp,cha+22jctc,Ditte2023,Ditte2024} (See Section \ref{ssec:jas}).

\subsection{Optimization methods}\label{ssec:opme}  

Within the framework of VMC in {\sffamily QMeCha} we have implemented the Stochastic Reconfiguration (SR) optimization scheme, introduced by S. Sorella\cite{sor+01prb,sor+05prb}, that does not require the computation of the local energy derivatives, and that, although usually converging more slowly than the Linear Method\cite{umr+07prl,tou+07jcp}, is more stable for a large set of parameters, and has much faster single step iterations\cite{bec+17}.

In SR, the variation of the parameters' vector $\boldsymbol{\alpha}$ is written as the equation:
\begin{equation}
\boldsymbol{\alpha}'=\boldsymbol{\alpha}+ \Delta \textbf{S}^{-1} \textbf{f}_{\boldsymbol{\alpha}},    
\end{equation}
where $\Delta$ is a constant damping parameter used to regularize the optimization\cite{sor+05prb}, reducing instabilities, $\textbf{f}_{\boldsymbol{\alpha}}=-\frac{\partial }{\partial \boldsymbol{\alpha}}\text{E}\left[\Psi_T(\boldsymbol{\alpha})\right]$ is the vector of the generalized forces that are computed as
\begin{equation}
f_\alpha = 
- 2 
\left \lbrace 
\left \langle
 \text{E}_{l} \mathcal{O}_\alpha
 \right \rangle_\Pi
- 
\left \langle
\text{E}_{l} 
\right \rangle_\Pi
\left \langle
\mathcal{O}_\alpha
\right \rangle_\Pi
\right \rbrace ,
\label{eq:f_a}
\end{equation}
and $\textbf{S}^{-1}$ is the inverse of the square symmetric matrix that defines the metric space of the parameters\cite{sor+05prb,bec+17}, defined as the covariance matrix 
\begin{equation}
\textbf{S}_{kl}=
\left \langle
 \mathcal{O}_{\alpha_k} \mathcal{O}_{\alpha_l}
 \right \rangle_\Pi
 -
 \left \langle
 \mathcal{O}_{\alpha_k} 
 \right \rangle_\Pi
 \left \langle
 \mathcal{O}_{\alpha_l} 
 \right \rangle_\Pi ,
 \label{eq:S_SR}
\end{equation}
of the derivatives of the logarithm of the trial wavefunction with respect to each parameter, \textit{i.e.} $\mathcal{O}_{\alpha_k}=\frac{\partial}{\partial \alpha_k}\ln [ \Psi_T(\boldsymbol{\alpha})]$. 

Here we must add that within the code, the $\textbf{S}$ matrices are never explicitly computed since, the linear system of equations
\begin{equation}
\textbf{S} \delta \boldsymbol{\alpha} = \textbf{f}_{\boldsymbol{\alpha}}
\end{equation}
can be computed via a conjugate gradient (CG) algorithm with parallel and distributed, nearly independent, computational tasks.
Since statistical noise can deteriorate the stability of the SR optimization, due to the fact that the eigenvalues of the covariance matrices can be very small, amplifying the noise in the stochastic forces, here we apply a constant positive shift of the diagonal matrix elements to regularize the matrix, as previously suggested in refs. \citenum{bec+17,sor+07jcp}, so that $\textbf{S}'=\textbf{S}+\epsilon \textbf{I}$ with $\epsilon \in [0.0001,0.01]$ depending on the different scales of the parameters\cite{bec+17,sor+07jcp}. Furthermore, in order to improve the convergence of the CG method, the $\textbf{S}$ matrices are preconditioned according to the procedure reported in ref. \citenum{bec+17}.
This optimization procedure is also implemented within the framework of the correlated sampling technique. 
In this case, the code measures the overlap between the wavefunction used for the initial sampling and the wavefunction at a given step and decides based on a threshold when to automatically resample if needed. 

\subsection{Diffusion Monte Carlo}\label{ssec:dmc}

In {\sffamily QMeCha} we have implemented a fixed-node diffusion Monte Carlo algorithm inspired by the one first published by Umrigar et al. in ref. \citenum{umr+93jcp}.

Since the systems described through the code are usually fermionic particles, as the QDO systems are composed of distinguishable bosonic particles, the approximation introduced to solve the sign problem is the Fixed-Node (FN) approximation~\cite{fou+01rmp,*kal+08ch8,*bec+17}. 

The fixed node procedure in {\sffamily QMeCha} differs from that in ref. \citenum{umr+93jcp} by a set of more modern features that have been introduced to improve the estimation of energy difference with pseudopotential calculations. 
In particular, to integrate the pseudopotential component, determinant locality approximation (DLA)\cite{zen+19jcp} is used to improve stability and reproducibility of the DMC energies independently on the type of Jastrow\cite{dellapia+25jcp}, with respect to locality approximation (LA)\cite{mit+91jcp}.
T-moves\cite{cas+10jcp} and determinant T-moves\cite{dellapia+25jcp}, that are more stable but computationally more expensive approaches have still to be implemented in the code, together with Lattice Regularized diffusion Monte Carlo (LRDMC)\cite{lrdmc05,cas+10jcp}. \\ 
Second, following the work of Anderson \textit{et al.} \cite{and+21jcp,and+24jcp} in order to reduce the size-consistency error of the DMC calculations\cite{zen+16prb}, a single particle energy cut-off algorithm has been introduced, removing the original dependency of the branching factor on the \textit{effective} time step\cite{and+24jcp} due to the fact that this increases the size-consistency error. 

The efficiency of the algorithm is shown in a code comparison investigation presented in ref. \citenum{dellapia+25jcp}.

\begin{figure*}[t!]
\centering
\includegraphics[width=0.8\textwidth]{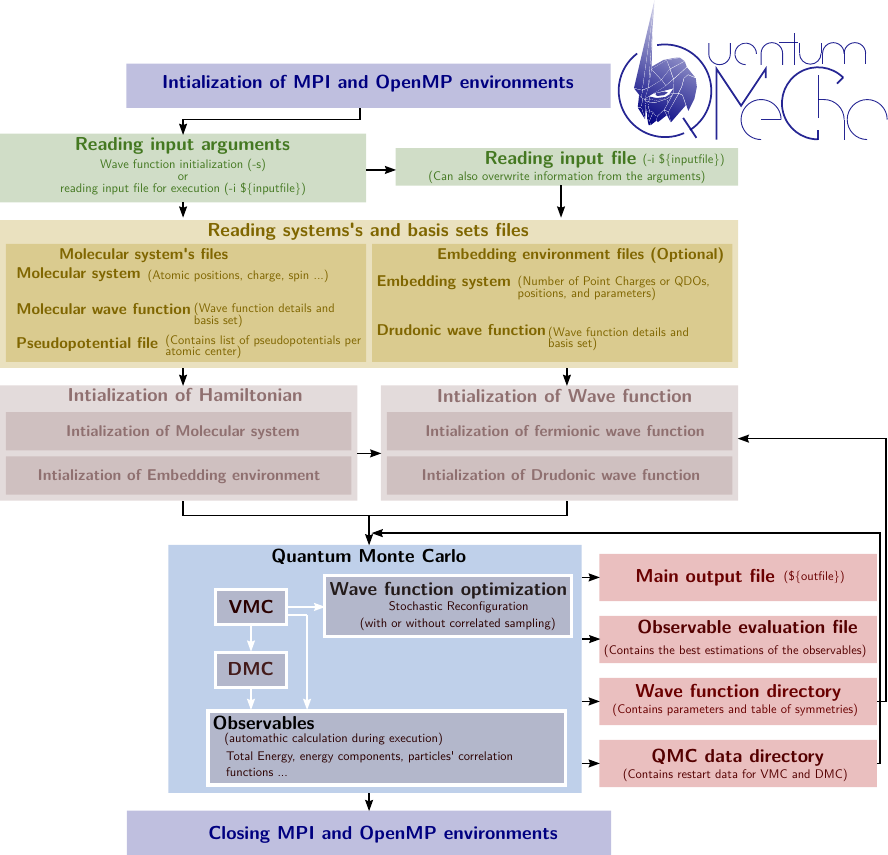}
\caption{
Schematic representation of the main structure of the {\sffamily QMeCha} code.
}
\label{fig:fig7}
\end{figure*}

\section{Code Structure, Computational details and Efficiency}\label{sec:comp_eff}

The {\sffamily QMeCha} code is a modular tool written in Fortran 2008 with two levels of parallelization achieved with both Message Passing Interface (MPI)\cite{mpi} and shared-memory Open Multi-Processing (OpenMP)\cite{openmp}. 

The code is available under Creative Commons Attribution-NonCommercial-NoDerivatives (CC BY-NC-ND) license through the GitHub repository {\ttfamily github.com/QMeCha}.
The GitHub repository of {\sffamily QMeCha} includes a repository of tools for the initialization of the wavefunction files {\ttfamily github.com/QMeCha/QMeCha\_tools} and the open version of the code {\ttfamily github.com/QMeCha/QMeCha\_code}. 

\begin{figure}[t!]
\centering
\includegraphics[width=\columnwidth]{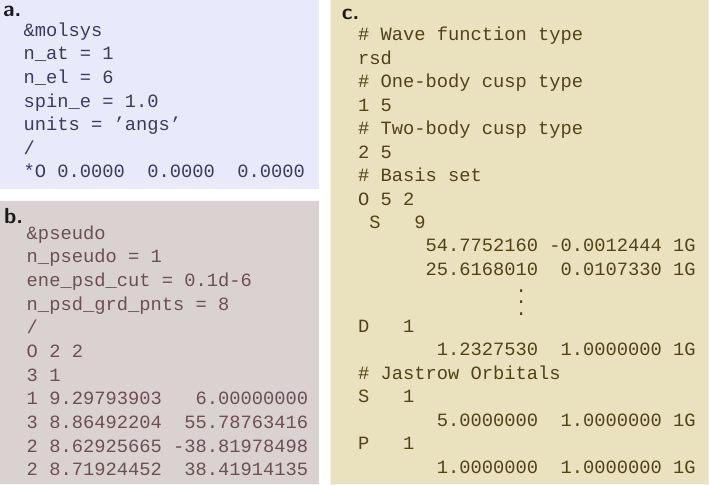}
\caption{
Main input files of the QMeCha code. \textbf{a.} Molecular system file. \textbf{b.} pseudopotential file (using the ccECP pseudopotential for Oxygen\cite{tra+05jcp1,*tra+14jctc,*tra+15jcp,*kro+16prb,*tra+17jcp,*tra+05jcp2,*ben+17jcp,*ben+18jcp,*ann+18jcp,*bur+07jcp,*bur+08jcp}). \textbf{c.} Wave function specifications, that include the type of wavefunction (RSD), the type of one- and two- body cusp functions (eq. \ref{eq:fun_csp}) and the cc-pVDZ basis set (truncated) for Oxygen, and the atomic Jastrow basis set. 
}
\label{fig:figinputs}
\end{figure}

The general structure of the code is briefly represented in Fig. \ref{fig:fig7}. 
The first step is the initialization of the MPI/OpenMP environment.
Afterwards, the code takes as input a series of arguments that are used to specify the input files that include the systems' properties, such as charge, spin, and atomic coordinates, and the basis set files. 
The input files for the fermionic system and the embedding environment are separated. 
These input files can also be specified in the main input file that includes the details of the QMC runs that have to be executed -- that is indicated as {\ttfamily \$\{inputfile\}} in Fig. \ref{fig:fig7} -- overwriting the information from the input arguments. 
A simple example of the main input files of {\sffamily QMeCha} can be found in Fig. \ref{fig:figinputs} for the case of the Oxygen atom described with the ccECP pseudopotential\cite{tra+05jcp1,*tra+14jctc,*tra+15jcp,*kro+16prb,*tra+17jcp,*tra+05jcp2,*ben+17jcp,*ben+18jcp,*ann+18jcp,*bur+07jcp,*bur+08jcp}.

If the wavefunction is first initialized through the \texttt{-s} argument, the molecular coefficients of the determinant parts can also be imported by reading the outputs of quantum Chemistry codes such as Orca\cite{ORCA} using the tools in the {\ttfamily github.com/QMeCha/QMeCha\_tools} repository.

Within the code, after the initializations of the molecular and embedding systems and of the basis sets and wavefunctions, the code calls the module that executes the VMC, DMC or wavefunction optimization procedures. 
Each molecular system property, each QMC method, and each wavefunction component are handled by separate {\ttfamily Fortran modules} progressively linked together in order to guarantee a simple way to progressively expand each main element of the code, without conflicts. 

This modularity is also important for future GPU acceleration, which will require to rewrite parts of the code in the appropriate programming language. 
In fact, although GPUs as accelerators have been used in computational science since the early 2000s, a portable and efficient protocol for all types of architectures is still missing -- despite the introduction of GPU interface in OpenMP.
Thus, an efficient porting of {\sffamily QMeCha} for different graphic cards must take into account compatibility of programming languages, libraries and compilers and the availability of the support for multi-GPU computing.

\begin{figure}[t]
\centering
\includegraphics[width=\columnwidth]{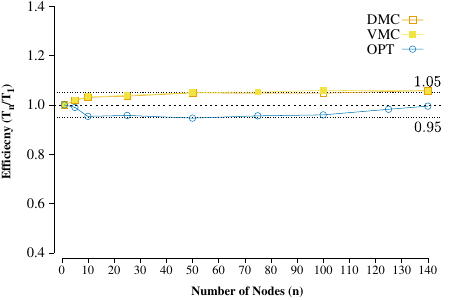}
\caption{Weak scaling tests for the system of 30 water molecules all described as electronic systems. 
The original geometry is taken from ref. \citenum{rak+19jcp} (\href{https://sites.uw.edu/wdbase/database-of-water-clusters/}{sites.uw.edu/wdbase}) and is shown in Fig. \ref{fig:fig1}a. }
\label{fig:fig5}
\end{figure}

Although the main purpose of {\sffamily QMeCha} up to now has been the creation of an easily modifiable platform, the intrinsic parallelization of the QMC algorithms guarantees near perfect weak scaling of the algorithms (see Fig. \ref{fig:fig5}) with respect to the number of nodes.

The tests reported in Fig. \ref{fig:fig5} are obtained from the system of 30 water molecules\cite{rak+19jcp} (\href{https://sites.uw.edu/wdbase/database-of-water-clusters/}{sites.uw.edu/wdbase}) using ccECP pseudopotentials\cite{ben+17jcp} with 240 electrons.
They have been run through the EuroHPC ecosystem on the MeluXina HPC of which each node includes two AMD EPYC 7H12 64-core processors with a total number of 128 physical cores and 256 threads.

For this set of tests, multi-threading has not been used, and thus each MC calculation has been run on 8 MPI tasks with 16 threads, which loop over the 16 Monte Carlo walkers per task. 
We can see that the latency deriving from the MPI communications is basically negligible also for DMC, for which a fixed population\cite{sor+00prb} algorithm, that requires more inter-MPI task communications, was used. 

\section{Applications}\label{sec:appl}

{\sffamily QMeCha} has been used in a wide range of applications. 
In particular, it was designed to compute interaction energies of large molecules interacting via van der Waals (vdW) interactions, which require a high level of accuracy. 
It has been also applied to study electron-positron metastable interactions in molecules, and to characterize changes in molecular bonding and excitation in embedding environments built from point charges and QDOs. 

In the next sections, we will describe these three main classes of applications. 

\begin{figure*}[t!]
\centering
\includegraphics[width=1.0\textwidth]{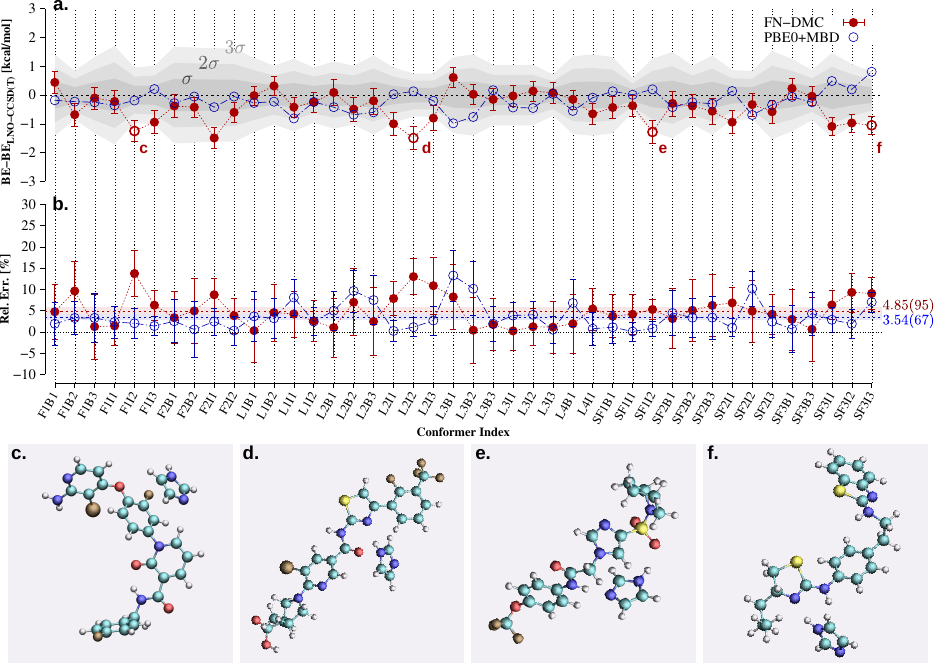}
\caption{\textbf{a.} Difference between the binding energies (BE) of the dimers computed with FN-DMC and DFT (PBE0+MBD) and the LNO-CCSD(T) calculations. 
For the LNO-CCSD(T) calculations we also report the estimated uncertainty that includes the basis set extrapolation and the error due to the orbital localization (see ref. \citenum{Puleva2025} for further details), indicated with the different $\sigma,2\sigma,3\sigma$ curves. These are only used as a guide for the eye to underline the dimers for which there is the largest discrepancy between FN-DMC and LNO-CCSD(T).
\textbf{b.} Relative error (in percentage) of the FN-DMC and DFT (PBE0+MBD) calculations defined with respect to the LNO-CCSD(T) ones as $| (\text{BE}-\text{BE}_{\text{LNO-CCSD(T)}} )/ \text{BE}_{\text{LNO-CCSD(T)}}|$. 
In this case the errors on the FN-DMC and PBE0+MBD calculations are obtained through the propagation of error that also includes the uncertainty in the LNO-CCSD(T) results. 
On the left we report the average relative absolute errors for FN-DMC and PBE0+MBD respectively in blue and red.
In panels \textbf{c}, \textbf{d}, \textbf{e}, \textbf{f} we display the structures of the conformers F1I2, L2I2, SF1I2 and SF3I3 respectively, that show the larges discrepancies between FN-DMC and LNO-CCSD(T) results, as shown in panel \textbf{a}.  Data from M. Puleva, L. Medrano Sandonas, B. D. Lőrincz, J. Charry, D. M. Rogers, P. R. Nagy and A. Tkatchenko
Nat. Communications \textbf{16}, 8583 (2025). ; licensed under a Creative Commons Attribution (CC BY) license.\cite{Puleva2025,QUIDrep}.}
\label{fig:fig3}
\end{figure*}

\subsection{Van der Waals interactions in macro-molecules}

In recent decades, the challenge of modeling increasingly large and complex molecules and materials has drawn attention to the role of vdW interactions in influencing the electronic, structural, spectroscopic, kinetic, and mechanical properties of biological and condensed matter phenomena\cite{rez+16cr,Tkatchenko2017rev,Hirschfelder}.

In order to model such subtle interactions, accurate quantum chemistry methods are usually required, at least to provide reference calculations that can be used to develop accurate effective or coarse grained approaches to be applied in macro-molecular systems\cite{rez+16cr}. 

The two reference methods that are usually used to provide accurate estimates for small and medium size systems\cite{rez+16cr,jurevcka2006benchmark,Dubecky2016,s66,s66x8,nenci,AlHamdani2021,Benali2014,shi+25jcp} are FN-DMC\cite{Foulkes2001,bec+17} and CCSD(T)\cite{bartlett+07rmp}.
On one hand, CCSD(T)\cite{bartlett+07rmp} is regarded as the \textit{`gold standard'} method of quantum Chemistry able to converge the dynamical correlation for energy and force estimations for a large variety of chemical compounds. 
On the other hand, also FN-DMC is expected to give similar or even more accurate results, since the method is able to recover all the dynamical electronic correlation responsible for such bindings.

Interestingly enough, in many recent works a significant and consistent discrepancy has been observed between the estimations of the binding energies provided through CCSD(T) and those obtained by FN-DMC on a single Slater determinant obtained from HF or DFT molecular orbitals.\cite{rez+15pccp,AlHamdani2021,Benali2014,shi+25jcp,nakano+25jctc}.
These recent studies highlight the tendency of FN-DMC to overestimate hydrogen bonds\cite{shi+25jcp,lam+25arxiv} and to sometimes underestimate dispersion bonding with respect to CCSD(T) -- this is not the case for example for the water-methane dimer\cite{dellapia+25jcp}. 
Thus, it is still debated which method yields more accurate results for a particular class of systems.\cite{sch+25nc}

Clearly, some of this discrepancy stems from the inconsistency in the description of the nodal structures of the vdW-interacting fragments with those in the non-interacting limit that, in some cases, such as the water dimer\cite{ste+08jctc,gur+07jcp}, require multi-determinantal approaches or backflow transformations to converge.
However, as also shown in a recent work by Shi \textit{et al.} \cite{shi+25jcp}, for the dispersive interacting systems of the S66\cite{s66,s66x8} dataset the FN-DMC results have only a mean average deviation of 0.09 kcal/mol with respect to the CCSD(cT)\cite{sch+25nc} calculations, that differ from CCSD(T) only on the improved approximation to the triple particle-hole excitation amplitudes. 
Furthermore, within these energy differences the convergence of CC with respect to the order of explicit and perturbative excitations included has been shown to not converge monotonically\cite{lam+25jcp}. 
Therefore, for these small interaction energies that become crucial in progressively large molecular systems, the puzzle of the differences between FN-DMC and CCSD(T) still remains unsolved, calling for the improvement of the state-of-the-art of both CC and FN-DMC methods.   
 
To shed some light on this puzzle, recently {\sffamily QMeCha} was used to obtain FN-DMC reference calculations, compared to LNO-CCSD(T)\cite{nagy+17jcp,LocalCC3,LocalCC4,MRCC}, for a set of molecular dimers each containing hundreds of electrons (see Fig. \ref{fig:fig3}) in ref. \citenum{Puleva2025}. 

The purpose of such reference calculations was to support the creation of a new dataset of molecular dimers interacting through vdW, \textit{i.e.} the ``Quantum Interacting Dimer'' (QUID)\cite{Puleva2025,QUIDrep}, built from the interaction of one macromolecule, taken from the Aquamarine (AQM)\cite{aquamarine} dataset, with benzene (C$_6$H$_6$) or imidazole (C$_3$H$_4$N$_2$).

The interactions between the chosen subset of the AQM molecules and the two small monomers have the property of including multiple vdW interaction types within each dimer, such as $\pi$-$\pi$ stacking, hydrogen bonding, electrostatics, and dispersion interactions. 

In Fig. \ref{fig:fig3}a, we have summarized the resulting differences between the binding energies (BE) computed with FN-DMC and PBE0+MBD with respect to the LNO-CCSD(T) results.
The conformers in the QUID dataset are originally catalogued through a string in which the first two letters indicate the main structural property of the macro-molecule, \textit{i.e.} folded (F), linear (L), semi-folded (SF), the first number indicates the number of the conformer, the letters I and B indicate the interaction with imidazole and benzene respectively and the last number indicates the binding site. 
Further details can be found in ref. \citenum{Puleva2025}.
In this panel, we also display the uncertainty on the LNO-CCSD(T) results that come from the localization and basis set extrapolation errors.
In Fig. \ref{fig:fig3}b, we report the relative error $\left | (\text{BE}-\text{BE}_{\text{LNO-CCSD(T)}} )/ \text{BE}_{\text{LNO-CCSD(T)}} \right |$ in percentage on the binding energies. Here, the estimated uncertainties on the LNO-CCSD(T) extrapolated results is propagated on both the FN-DMC and the PBE0+MBD reported errors. 

From these results, it is evident that the biggest discrepancies can be found in those dimers involving the macromolecule interacting with imidazole and presenting hydrogen bonds. 
In particular, in Fig. \ref{fig:fig3}a we highlight four dimers that are displayed in Figs. \ref{fig:fig3}c,\ref{fig:fig3}d,\ref{fig:fig3}e,\ref{fig:fig3}f.

In all these conformers, there appears to be a hydrogen bond between the nitrogen atom of one monomer and the hydrogen of the other, sometimes the donor is imidazole (Figs. \ref{fig:fig3}d,\ref{fig:fig3}f) and sometimes the macro-molecule  (Figs. \ref{fig:fig3}c,\ref{fig:fig3}e).
Clearly, it is impossible to exclude that this effect comes from the inconsistency between the description of the nodal structure of the bonded dimer and of the non-interacting one, thus the question of how the convergence of the nodal surface affects the accuracy of the descriptions of such interactions still remains an open question that deserves further investigation.

In general, these results confirm what was also observed in refs. \citenum{AlHamdani2021,Benali2014,shi+25jcp} that FN-DMC with a Slater determinant wavefunction tends to overestimate the binding energies of hydrogen-bonded dimers by about 0.1 kcal/mol, while the results obtained for dispersive interacting systems can achieve significant accuracy, and can clearly be used as reference calculation for the construction of less expensive computational methods.

\begingroup
\squeezetable
\begin{table*}[t!]
\scriptsize
\caption{Non-relativistic total energies (in Hartree) of atoms interacting with a positron ($e^{+}$) or with positronium (Ps). In parenthesis, we report the symmetry state of the electrons. All calculations are done with the SD or AGP wavefunction and the electron-positron wavefunction in eq. \ref{eq:EPO}. 
SP indicated the single-pairing function that corresponds to one anti-symmetrized explicitly correlated pairing function from ref. \citenum{Mella1999}, while MP is the linear combination of SP functions. Data partially taken from J. A. Charry Martinez, M. Barborini and A. Tkatchenko J. Chem. Theory Comput. \textbf{18}, 4, 2267–2280 (2022); licensed under a Creative Commons Attribution (CC BY) license.\cite{cha+22jctc}.}\label{tab:tab1_positron_atoms}
\begin{ruledtabular}
\begin{tabular}{lcccccccc}
& {e$^{+}$Li($^2$S)} & {e$^{+}$Be($^1$S)} & {PsH($^1$S)}    & {PsLi($^1$S)} & PsB($^3$S) & {PsC($^4$S)}  &  {PsO($^2$P)}  & {PsF($^1$S)}   \bigstrut \\ 
~\\[-2mm]
VMC SP \cite{Mella1999} & -7.52510(10)      &                   & -0.786200(10)     & \\
VMC MP \cite{Mella1999} & -7.530180(10)     &                   & -0.788230(10)     & -7.726160(80) \\
VMC \cite{bre+98jcp}    &                   &                   &                   & -7.498200(30) & -24.765(2)        &  -38.0030(20)  & -75.1450(30)     & -99.9960(30)	 \\[1mm]
VMC SD   & & & & & -24.84097(13) \\
VMC AGP  & -7.52566(80)      & -14.66386(18)     & -0.786416(33)     & -7.723921(87) & -24.846154(81)    &  -38.06800(39) & -75.28366(53)    & -100.02490(58) \\[1mm]
DMC SP \cite{Mella1999} & -7.531650(80)     &                   & -0.789160(30)	    & \\
DMC MP \cite{Mella1999} & -7.532290(20)     &                   & -0.789150(40)     & -7.739529(60) \\
DMC \cite{bre+98jcp}    &                   &                   &                   & -7.737600(40) & -24.875(1)        & -38.09590(60) & -75.31770(50)     & -100.07190(80) \\[1mm]	
DMC SD & & & & & -24.87563(82)       \\
DMC AGP   & -7.53094(23)      & -14.66931(36)     & -0.7891191(31)    & -7.73804(41)  & -24.87819(37)     & -38.09795(57) & -75.32969(63)     & -100.07435(15) \\[1mm]
CI         &              &              & -0.78874(60)\footnotemark[1]& & -24.83056\footnotemark[2]  & -38.05362\footnotemark[2] & -75.28127\footnotemark[2] &  -100.001817\footnotemark[3] \\
SVM        & -7.532323\cite{Ryzhikh1998}   & -14.669042\cite{Ryzhikh1998}  & -0.789196\cite{Mitroy2006}   & -7.740208\cite{Mitroy2001}\\
Hylleras\cite{Yan1999}                &                   &                   & -0.7891967147(42) \\
\end{tabular}
\footnotetext[1]{FCI extrapolation from ref. \citenum{cha+18ang}.}
\footnotetext[2]{FCI limit with higher momentum corrections from ref. \citenum{Saito2006}.}
\footnotetext[3]{MRCI calculation from ref. \citenum{Saito2005}.}
\end{ruledtabular}
\end{table*}
\endgroup

\subsection{Interacting Electron-Positron systems}

The {\sffamily QMeCha} code was developed to study explicitly correlated wavefunctions for many-particle systems.
In this context, an intriguing set of applications that extend beyond the complexity of many-electron correlations is the study of meta-stable states that form between molecules and positrons \cite{Natisin2017,Mitroy2002,Gribakin2010}. 

The positron is the anti-particle of the electron, with which it shares the same mass and spin statistics while having opposite charge and lepton number\cite{and+33pr}. 
Experimentally, positrons are often used as probes in spectroscopic techniques applied in chemistry, biology, and materials science\cite{wahl2002principles,jean2003principles,RevModPhys.85.1583}. These techniques are based on the detection and analysis of gamma rays produced from the electron-positron annihilation process.

Moreover, since the early 2000s the methods to accumulate and manipulate positrons \cite{surko.rmp.2015} and positronium (Ps) \cite{Mills2019} at low energies have greatly advanced, allowing the production of dipositronium (Ps$_2$) \cite{Cassidy2007}, the development of positronium gamma-ray lasers \cite{Cassidy2018}, and the production of long-lived positronium beams to study gravitational interactions\cite{Amsler2019}.

At low energies, it has been experimentally observed through resonant annihilation\cite{Natisin2017,Mitroy2002,Gribakin2010} that before the annihilation process with electrons, positrons can form bound states with atoms and molecules, with a lifetime of about 10$^{-9}$ seconds, that is longer than some vibrational motions and thus able to interfere with chemical reactions.

These experimental findings have stimulated a wide range of theoretical studies\cite{Gribakin2010}, which have suggested various binding mechanisms between the positron and the atomic\cite{Mitroy2002,Cheng2012,Harabati2014} or molecular compounds\cite{Charry2018,Moncada2020,Goli2019,shu+20jcp,Bressanini2021.one,Bressanini2021.two}. 
Furthermore, it has been shown that positrons can act as a chemical mediator able to change the energy profiles of proton-transfer reactions in aminoacid compounds~\cite{Suzuki2019}.

The theoretical description of the ground state of these meta-stable states can be obtained through the solution of the time-independent Schr\"odinger equation with the general Hamiltonian described in eq. \ref{equ:ferm_ham}. 

If the correlation between electronic pairs in a molecular system is already a challenging problem, the presence of a positron introduces additional complexity because of the attraction with the electrons. 
The positron can, in fact, be seen as a very light nucleus that forms bound states with the electrons in the molecular system and is repelled by the nuclei, requiring expensive explicitly correlated methods\cite{str+01jcp,bub+04jcp,swa+12jcp,Ryzhikh1998,Zhang2007} that also consider the explicit dependency on the electron-positron inter-particle distances.

For this reason, QMC methods\cite{bre+98jcp,Schrader1998,Mella1999,bor+05appa,kit+09jcp,Kita2010,dru+11prl,Cassella2024,upa+24jctc} stand out as a computationally efficient alternative to deterministic approaches based on numerical integration, such as many-body perturbation theory (MBPT)\cite{Hofierka2021} and configuration interaction (CI) \cite{rey+19ijqc,Saito2003,Saito2005,Gribakin2010,coe_CPL_645_106_2016}.

Naturally, in order to efficiently describe these systems, QMC methods require the construction of a parametric trial wavefunction that includes the most important characteristics of the exact wavefunction, as done in previous studies\cite{bre+98jcp,schrader_RAiCC_2_163_1997, Schrader1998,Mella1999,bor+05appa,kit+09jcp,Kita2010,dru+11prl,yam+14epjd}.  

In particular, beyond the standard electronic correlation effects and cusp conditions, the wavefunctions must also include the attractive correlation effects between the electron-positron pairs, describing the cusp conditions and including the correct asymptotic behaviour as a function of the inter-particle distances\cite{bre+98pra,bre+98jcp,mel+00jcp_1,mel+00jcp_2,Bressanini2003,swa+12jcp,bro+17jpca}. 

In this first release of the {\sffamily QMeCha} code, the wavefunction implemented to describe positronic systems includes that of the ``any particle molecular orbital approach''\cite{rey+19ijqc,kit+09jcp,Kita2010}, and also the more efficient product of electronic wavefunction times that of the geminal positronic wavefunction\cite{bre+98jcp,swa+12jcp,bro+17jpca} (see Section \ref{ssec:eleposgem}).

These two wavefunctions have been used first in ref. \citenum{cha+22jctc} to compute the positron affinity and Ps binding energies for a set of atoms and the $e^+\cdot(\text{H}_2^{2-})$ molecular system.
All energies were compared against the most accurate reported values in the literature at the time, either from CI, analytical explicitly correlated methods, or VMC and DMC. 
As can be seen from the data summarized in Table \ref{tab:tab1_positron_atoms}, for the positronium-bound systems: PsB, PsC, PsO, and PsF, the VMC and DMC total energies achieved were lower than any previously reported calculations, establishing new benchmark values, while the PsH and PsLi results are remarkably competitive against larger multi-reference values. 
On the other hand, there is still room for improvement at VMC level for the weakly bound multireference systems such as $e^+\text{Li}$ and $e^+\text{Be}$.

\begin{figure}[t!]
\centering
\includegraphics[width=\columnwidth]{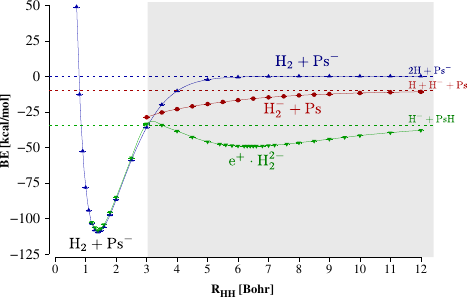}
\caption{Potential energy surfaces of $e^+\cdot(\text{H}_2^{2-})$, $2\text{H}+\text{Ps}^-$ and $\text{H}+\text{H}^-+\text{Ps}$ obtained at the DMC level and reported in ref. \citenum{cha+22jctc}. Partially adapted from J. A. Charry Martinez, M. Barborini and A. Tkatchenko J. Chem. Theory Comput. \textbf{18}, 4, 2267–2280 (2022); licensed under a Creative Commons Attribution (CC BY) license.
}
\label{fig:fig9}
\end{figure}

Regarding positronic molecules, the same methodology was tested for $e^+\cdot(\text{H}_2^{2-})$, a particular system characterized by an unique stabilising positron bond interaction between two repelling hydrogen anions \citenum{Charry2018, shu+20jcp, Bressanini2021.one}. 
The stability of $e^+\cdot(\text{H}_2^{2-})$ was a matter of debate due to presence of an apparent double-well minima, that at short distances corresponds to a highly delocalized unbound state of $\text{H}_2$ and Ps$^-$ \cite{Bressanini2021.one}, while at longer distances as displayed in the PES in Fig. \ref{fig:fig9}, shows a local stable positron bounded $e^+\cdot(\text{H}_2^{2-})$ structure. 
Compared to the most accurate DMC values \cite{Bressanini2021.one}, the AGP/EPO, and AGP/PMO wavefunction shown only in ref. \citenum{cha+22jctc}, demonstrated the ability to describe both minima, while previous CI or HF+J VMC approaches could not clearly capture the dissociated Ps$^-$ state\cite{Bressanini2021.one,cha+22jctc}.

The overall results validate the reliability and transferability of the new correlated wavefunctions and their robust variational optimisation for describing electron-positron interactions, confirming the approach as a promising general tool for larger electron-positron systems.

\begin{figure}[ht!]
\centering
\includegraphics[width=\columnwidth]{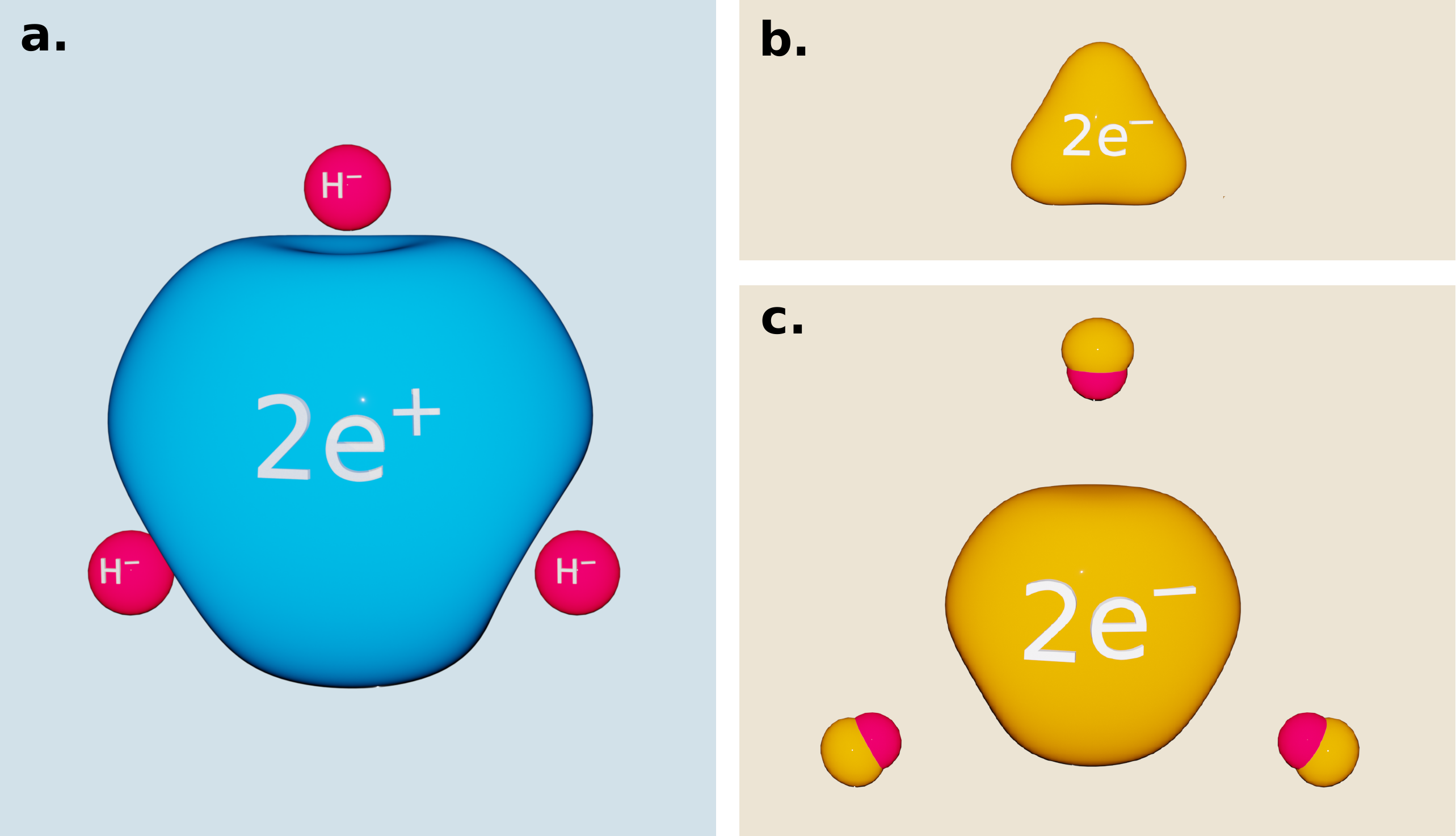}
\caption{\textbf{a.} Positronic (blue) and electronic (red) densities for $2e^+\cdot(\text{H}_3^{3-})$. \textbf{b.} Electronic (orange) density in $\text{H}_3^{+}$. \textbf{c.} Core (red) and valence (orange) electronic densities in $\text{Li}_3^{+}$.
Adapted from J. Charry, F. Moncada, M. Barborini, L. Pedraza-González, M. T. do N. Varella, A. Tkatchenko and A. Reyes, Chem. Sci. \textbf{13}, 13795-13802 (2022); licensed under a Creative Commons Attribution (CC BY NC) license.\cite{cha+22cs}.
}
\label{fig:fig8}
\end{figure}

In a subsequent investigation\cite{cha+22cs}, {\sffamily QMeCha} was employed to study the energetic stability of a system containing two positrons and three hydride 
anions in $2e^+\cdot(\text{H}_3^{3-})$, confirming through QMC calculations, the existence of a local stable equilibrium geometry with D$_{3h}$ symmetry. 
Its bonding properties and positronic densities were compared with those of the purely electronic analogues $\text{H}_3^{+}$ and $\text{Li}_3^{+}$ systems. 
Through such analysis, striking similarities between the two-positron compound and the trilithium cation were found, pointing out the formation of three-center two-positron bond, analogous to the well-established three-center two-electron counterparts as depicted in Fig \ref{fig:fig8}, extending then the concept of positron bonded molecules as seen for the $e^+\cdot(\text{H}_2^{2-})$.
Furthermore, in a recent work by Roldan \textit{et al.}\cite{rol+25cs} {\sffamily QMeCha} has been used to compute the potential energy surface of the $e^+\cdot(\text{Be}_2)$ molecules.

Further implementations are ongoing to better develop the positronic wavefunctions in a general, stable, and consistent format, which will be the topic of future releases of {\sffamily QMeCha}.

\subsection{Embedding approaches for quantum Monte Carlo}

Embedding methods, including those combining quantum mechanics and molecular mechanics (QM/MM)~\cite{Fersht2013,Sun2016,Jones2020,Khare2007,Torras2009,Zimmerman2011,Nasluzov2001,Nasluzov2003,Brandle1998,Guareschi2016,Senn2009,Acevedo2010,Huang2017,Alves2015,Silva2015,Bao2019,Marin2019,Robertazzi2006,Gkionis2009,Kupfer2014,Osoegawa2019,pia+15jcpb,hug+20jcc}, those merging different levels of quantum mechanical approaches (QM/QM)~\cite{Kotliar2006,Lan2015,Knizia2012,Wesolowski2015,Libisch2014,Manby2012,Katin2017,Friedrich2007,Seth2002,Vogiatzis2015,Huo2016,Lee2019,Coughtrie2018,Neugebauer2005,Jacob2006,Exner2012,Exner2004,Bockstedte2018,Lau2021,Schafer2021,Ma2021,Petocchi2020,Yeh2021,Nowadnick2015,Chen2022,Nusspickel2022} and those using implicit solvents such as the Polarized Continuum Model (PCM)~\cite{men+12cms}, the Surface Generalized Born model (SGB)~\cite{gos+98jpcb,rom+04jpca}, the
Conductor-like Screening Model (COSMO)~\cite{kla+95jpc}, and the Reference Interaction Site Model (RISM)~\cite{gan+22jpca,ima+24jcp},
are crucial tools to describe fundamental phenomena in molecular and materials science.

In fact, these methods are essential to model many important chemical and physical phenomena that extend over multiple energetic, spatial, and temporal scales\cite{Sun2016,Tkatchenko2017rev,Jones2020}.
To treat these different scales, embedding techniques simultaneously tackle partitioned subsystems with different levels of computational accuracy, paving the way to a feasible description of such phenomena.

The importance of developing such techniques within the framework of QMC methods\cite{Filippi2012,Valsson2013,Ditte2023,Ditte2025} lies in the possibility of working with any articulated parametric wavefunction and with multiple environments' parametrizations, allowing the exploration of new approaches to include explicit correlation between the environment and the molecular subsystem described through the electronic Schr\"odinger equation.
Furthermore, QMC can be used to obtain accurate reference calculations supporting the development of computationally cheaper approaches. 
\begin{figure}[t!]
\centering
\includegraphics[width=1.0\columnwidth]{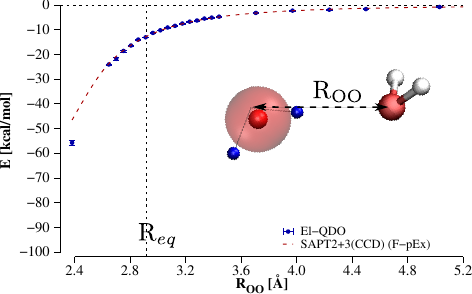}
\caption{ Interaction energies of the El-QDO water dimer as a function of the distance between the two subsystems. The results are compared to the SAPT2 + 3(CCD) energy components. The vertical dashed line is the equilibrium geometry of the water dimer.
The reported F-pEx curve is obtained by removing from the total SAPT energy all the pure exchange contributions\cite{Ditte2025}.
Data adapted from M. Ditte, M. Barborini and A. Tkatchenko, J. Chem. Theory Comput. \textbf{21}, 9, 4466–4480 (2025); licensed under a Creative Commons Attribution (CC BY) license.\cite{Ditte2025}.}
\label{fig:fig4}
\end{figure}

For this purpose, in the {\sffamily QMeCha} code, we have introduced a new quantum embedding approach, called El-QDO\cite{Ditte2023,Ditte2024,Ditte2025}, in which a molecular system described through the electronic Hamiltonian is embedded in an environment of Coulomb-coupled charged harmonic oscillators, \textit{i.e.} quantum Drude oscillators (QDOs)~\cite{london+36tfs, Ditte2024, Martyna2013} and point charges\cite{MartynaPRL2013,MartynaMolPhys2013,sok+15pnas}. 
This new joint framework also has the advantage of overcoming difficulties and limitations in the construction of embedding methods between QMC and polarizable FF~\cite{Guareschi2016} or between QMC and other first-principles approaches, such as DFT~\cite{var+14jcp}.

QDOs are parametrized to describe the response properties of real atomic environments, reducing the degrees of freedom with respect to the description obtained through the full electronic Hamiltonian~\cite{Martyna2006,Martyna2009,Martyna2013,Tkatchenko2009,Tkatchenko2012,Tkatchenko2017rev,Vaccarelli2021,Goger2023}. 
In particular, QDOs interacting via dipole-dipole potential, serve as the foundation of the many-body dispersion (MBD) method~\cite{Tkatchenko2012,Ambrosetti2014,Hermann2020} used in the framework of DFT~\cite{Tkatchenko2017rev,Gray2024} to provide the dispersion contributions to the energy and atomic forces that are usually underestimated in traditional exchange-correlation functionals~\cite{Tkatchenko2017rev,Gray2024}.

Coulomb-interacting QDOs~\cite{Martyna2006,Martyna2009,MartynaPRL2013,MartynaMolPhys2013,Manby2016,sok+15pnas,Ditte2024}, in particular, have previously been employed as models of atomic systems, in combination with DMC and path integral Monte Carlo (PIMC) methods to investigate dispersion interactions in noble gas dimers~\cite{Martyna2013}, crystals~\cite{Martyna2009} and fluids~\cite{Martyna2006}, and to study the structural and dynamical properties of liquid water~\cite{MartynaPRL2013,sok+15pnas} and ice~\cite{Martyna2019,MartynaMolPhys2013}.
QDOs have also been utilized within a full configuration interaction (FCI) framework, where oscillator wavefunctions are expanded in Gaussian basis sets and applied to prototypical dispersion systems~\cite{Manby2016}. 
Further extensions include the construction of universal pairwise van der Waals potentials~\cite{Khabibrakhmanov2023}, as well as studies of dipole-bound anions by coupling QDOs to single electrons within perturbation theory~\cite{Wang2001,Sommerfeld2005}.

In the El-QDO embedding method\cite{Ditte2023,Ditte2025} the degrees of freedom of the electronic targeted subsystem and those of the drudons in the environment are described through the single comprehensive Hamiltonian in eq. \ref{eq:H^tot}, describing explicit many-body correlation effects between the electronic sub-system and the environment, with the exclusion of the short-range interactions that are not captured. 
All the degrees of freedom are described through a collective variational ansatz, in eq. \ref{eq:tot_wf}, that can be easily integrated in the VMC and DMC frameworks. 
\begin{figure}[t]
\centering
\includegraphics[width=\columnwidth]{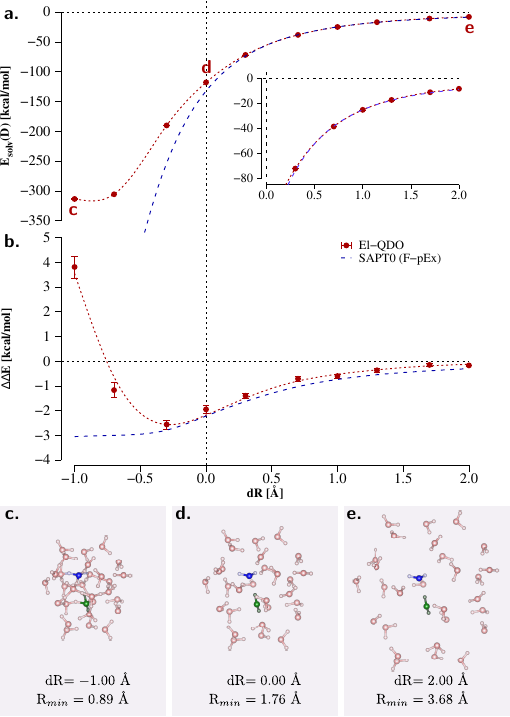}
\caption{
\textbf{a.} Solvation energy of the water dimer in the embedding cluster of 28 water molecules modelled by QDOs and point charges\cite{Martyna2019}. 
\textbf{b.} Variation of the water dimer's bond energy as a function of the interaction with the cluster of 28 water molecules.
In panels \textbf{c}, \textbf{d}, \textbf{e} we display different configurations of the embedding cluster of waters. 
$R_{min}$ correspond to the minimal distances between the atoms of the dimer and those of the environment. The original geometry for dR = 0.0 \AA \ is taken from ref. \citenum{rak+19jcp} (\href{https://sites.uw.edu/wdbase/database-of-water-clusters/}{sites.uw.edu/wdbase}). 
Data adapted from M. Ditte, M. Barborini and A. Tkatchenko, J. Chem. Theory Comput. \textbf{21}, 9, 4466–4480 (2025); licensed under a Creative Commons Attribution (CC BY) license.\cite{Ditte2025}.}
\label{fig:fig10}
\end{figure}

For hydrogen-bonded systems, such as the water dimer, we can see in Fig. \ref{fig:fig4} adapted from ref. \citenum{Ditte2025}, that the interaction energy between a QDO-modeled donor water molecule and a full electronic acceptor monomer\cite{Ditte2025} is a good approximation to a subset of energy components from Symmetry Adapted Perturbation theory SAPT2 + 3(CCD). 
The reported F-pEx curve is obtained by removing from the total SAPT energy all the pure exchange contributions\cite{Ditte2025}.
The removal of these exchange contributions comes from the fact that the El-QDO model is still not generalized to reproduce the short-range interaction limit between the environment and the molecular sub-system described at the electronic level. 
Following the study of the interaction energies of molecular and noble gas atoms\cite{Ditte2023,Ditte2025} the El-QDO approach was also used to study the solvation energies of benzene and water dimers, and the variations of their bond energies in cages of water molecules\cite{Ditte2023,Ditte2025}.
In Fig. \ref{fig:fig10} we report the solvation energies and bond energy variations for the water dimer as shown in ref. \citenum{Ditte2025} as a function of the cage distortion from equilibrium.
Here the solvation energy is defined, following ref. \citenum{Sirianni2022}, as the interaction energy of the molecular dimer with the cage $\text{E}_{solv}(\text{D})=\text{E}(\text{D}+\text{cage})-\text{E}(\text{D})-\text{E}(\text{cage})$, where $\text{E}(\text{D})$ and $\text{E}(\text{cage})$ are the energies of the isolated QDO cage and electronic molecular dimer, and $\text{E}(\text{D}+\text{cage})$ is the total energy of the interacting system.
The variation of the dimers bond energy,  $\Delta \Delta \text{E} = \text{E}_{solv}(\text{D}) - \text{E}_{solv}(\text{M}_1) - \text{E}_{solv}(\text{M}_2)$, is defined as the difference between the solvation energy of the dimer and those of the two monomers\cite{Sirianni2022}.

Both quantities are reported in Figs. \ref{fig:fig10}a and \ref{fig:fig10}b respectively, as a function of the contraction and expansion of the water cage of 28 molecules that surround the electronic water dimer.
The contraction and expansion of the cage is obtained using as reference point the geometric centre $\textbf{R}_C$ of the nuclei of the electronic subsystem. 
The vector of the positions of the oxygen atoms $\textbf{R}_O$ is modified by a scalar value dR according to the equation $\textbf{R}'_O=\textbf{R}_O(1+\frac{\text{dR}}{\sqrt{|\textbf{R}_O|^2}})$ along the vector connecting the atoms with $\textbf{R}_C$. 
The Hydrogen pairs of the water molecules are then translated according to shift of their Oxygen atom preserving the intramolecular angles and bonds so that the single molecular structures remain unchanged. 

In Figs. \ref{fig:fig10}c, \ref{fig:fig10}d and \ref{fig:fig10}e the structures of the water dimer in the cage of 28 water molecules are displayed for contractions and expansions of dR=[-1.0,0.0,2.0] where the value dR=0.0 \AA\ corresponds to the original MP2 equilibrium geometry from ref. \citenum{rak+19jcp}.
In these panels the values of $\text{R}_{min}$ correspond to the minimum distance between the QDOs in the environment and the atoms of electronic water dimer.
\begin{table}[t!]
\caption{
Relative runtimes of ortho-Benzyne (in singlet state), Benzene and the Benzene dimer computed in vacum (V) and in cages of four (4W) thirty (30W) or fifty (50W) water molecules. See ref. \citenum{Ditte2023}. 
El calculations are described at the full electronic level while the El-QDO DMC ones are done by substituting the water molecules with the QDO embedding environment.
Test calculations have been done with VMC calculations on 56 CPUs using 12 walkers per CPU and 20 bins with 100 steps per block. 
The time per single block is averaged over 20 blocks. 
Relative runtime is calculated as the ratio of time per block multiplied by the square of the ratio of the stochastic error $\sigma$, in order to take into account also the differences in the error bars between the cases in vacuum and in the embedding cage.
Data adapted from M. Ditte, M. Barborini, L. Medrano Sandonas, and A. Tkatchenko Phys. Rev. Lett. 131, 228001 (2023); licensed under a Creative Commons Attribution (CC) license.\cite{Ditte2023}
}\label{tab:runtime_dmc}
\begin{ruledtabular} 
\begin{tabular}{lccc}
&  Time per Block  & $\sigma$  & Runtime relative \\ 
&  [sec.] &  [E$_h$] & to vacuum   \\[2mm] 
    \multicolumn{4}{c}{{\it ortho-Benzyne (S)}}\\
    El (V)        &    6.36 &  0.80 & 1.00   \\
    El (4W)       &   25.10 &  1.40 & 11.99   \\
    El-QDO (4W)   &    6.45 &  0.80 & 1.02   \\
    El-QDO (30W)  &    6.78 &  0.87 & 1.25   \\[2mm] 
    \multicolumn{4}{c}{{\it Benzene}}\\
    El (V)        &    7.15 &  0.80 & 1.00   \\
    El-QDO (50W)  &    8.34 &  0.89 & 1.47   \\[2mm] 
    \multicolumn{4}{c}{{\it Benzene dimer}}\\
    El (V)        &    30.95 &  1.31 & 1.00   \\
    El-QDO (50W)  &    32.90 &  1.57 & 1.27   \\
\end{tabular}
\end{ruledtabular}
\end{table}

It is important to underline that the discrepancies between the DMC and SAPT0 (F-pEx) curves are due to the cut-off in the short range of the El-QDO calculations, and on the fact that the repulsive components are missing from the total potential to highlight the correspondence between El-QDO and SAPT0 (F-pEx) in the long range and around the equilibrium geometry\cite{Ditte2025}. 
Clearly, further investigations are still ongoing to improve the procedure and generalize the El-QDO approach.

Here, it is crucial to notice that, although the approach treats the drudons of the QDOs as quantum degrees of freedom in the systems (eq. \ref{eq:H^O}), the computational cost of the integration of the embedding environment is negligible with respect to the cost of the integration of the electronic degrees of freedom, as shown in Tab. \ref{tab:runtime_dmc}.  
In this table we report the computational cost of the DMC calculations for three sets of system, the first that contains one o-benzyne molecule and four water molecule, is described using both full electronic structure and substituting the four water molecules with 4 QDOs (El-QDO). 
The energy comparisons of the two approaches, reported in ref. \citenum{Ditte2023}, show perfect agreement between the two calculations, yet the computational cost of the El-QDO corresponds essentially to that of the single o-benzyne molecule in vacuum (For full comparisons see Fig. 4 of ref. \citenum{Ditte2023}).

This is due to the fact that while the wavefunction of the electronic system is usually represented by a determinant and its computational cost is of $\mathcal{O}(N_e^3)$, for the drudonic particles, that are distinguishable due to the asymmetry in the external potential\cite{Ditte2024}, the wavefunction can be written as a simple Hartree product whose computational cost is two orders of magnitude smaller.

Clearly, this advantage also holds beyond the framework of QMC, making the El-QDO method generalizable for other ab initio methods such as Configuration Interaction or Coupled Cluster opening the way to novel QM/QM approaches where the environment is treated through quantum drudonic degrees of freedom.

\section{Conclusions}\label{sec:concl}

We presented the first release of the {\sffamily QMeCha} QMC software -- a flexible code to study many-body fermionic systems in vacuum or embedded in semi-quantum polarizable force fields of point charges and QDOs.
In this review, after detailing the properties of the Hamiltonian, of the wavefunctions used to describe the various types of particles, and the main characteristics of the QMC methods used to integrate them, we have reported on three main sets of applications that have been enabled by the QMC methods implemented in {\sffamily QMeCha}. 

With the first application, related to vdW interactions\cite{Puleva2025} we have shown the accuracy of modern FN-DMC algorithms to describe these types of bonding, also compared to the golden standard of quantum chemistry, \textit{i.e.} CCSD(T). 
Thus, these results have confirmed the possibility of using QMC methods, and in particular FN-DMC, to construct accurate reference calculations on large molecular systems, which are crucial for the construction of modern flexible and portable Machine Learning (ML) and Deep Learning (DL) force fields\cite{slo+24jctc,pol+25cs_1,pol+25cs_2,Puleva2025}.

The second set of applications\cite{cha+22jctc,cha+22cs} has highlighted how QMC methods are paramount in the exploration of new positronic complexes, expanding our understanding of chemical bonds and positron-matter phenomena. 
The ability of QMC to integrate explicitly correlated wavefunctions, opens the way to the description of positronic molecules in terms of electron-positron orbitals that are constructed on the eigenfunctions of the bound states of positronium (Ps). These functional forms enable the modeling of these systems without a priori knowledge of their spatial location and angular momentum.
Moreover, the {\sffamily QMeCha} capability of optimizing a large set of variational parameters allows us to achieve the full relaxation of the correlated trial wavefunction, providing a better reference than the bare uncorrelated single Slater determinant electron-positron wavefunction, which is, in most cases, an incorrect unbound starting solution. 

In the third set of applications, we have briefly reviewed the El-QDOs embedding method\cite{Ditte2023,Ditte2024,Ditte2025} in which a molecular subsystem described through the electronic Hamiltonian is embedded in a bath of QDOs and point charges\cite{Martyna2006,Martyna2009,Martyna2013,Martyna2019}. 
The El-QDO method has been shown to be able to recover the dynamical correlation between the embedding environment and the electronic molecular systems, which is crucial for describing how the environment affects, solvation, electronic excitations, and bond energies.

The development of the {\sffamily QMeCha} code is still in its early stages and much has yet to be improved for all sets of applications presented in this review.
For example, to compute large vdW systems, more efficient FN-DMC methods and ans\"atze can be developed to reduce the time-step error and ensure convergence. 
For electron-positron systems, further generalization of the fermionic wavefunction and of the Jastrow factor is already in development.
Finally, a generalization of the El-QDO method is being designed to tackle more complex embedding cases and to efficiently include dynamical processes. 

Clearly all this is stimulated by the fact that QMC is still the method of choice for the exploration of novel Hamiltonian models and ans\"atze. 
In fact, for example, in the last few years the development of transcorrelated methods\cite{hau+25jcp} and the deterministically optimized Jastrow factor \cite{fil+25jcp} have been made possible through previous research on QMC techniques and variational ans\"atze.

In conclusion, within the framework of the QMC methods, {\sffamily QMeCha} serves as a versatile platform for experimentation and for the construction of new models to describe many-body correlations between thousands of particles of fermionic and bosonic symmetry.\\
In the future, beyond the applications described above, the code will be extended to include novel \textit{backflow} transformations\cite{dru+06jcp} like those constructed through neural networks\cite{paulinet,pfa+20prr,pfa+24sci}, and to efficiently compute forces through VMC and DMC methods\cite{rhi+22jctc,nak+22jcp,slo+24jctc}, further extending the applicability of QMC methods as important reference tools for the description of molecular and solid-state systems.  

\begin{acknowledgements}
MB is deeply in debt with Prof. Sandro Sorella, who has been a mentor to many in the fields of strongly correlated systems and quantum Monte Carlo methods, among others. 
MB first met Prof. Sorella in 2009 as a co-supervisor for his master thesis in physics, and since then he became an essential inspiration and support for his scientific career. 
Sandro Sorella's brilliant mind and sincere kindness are greatly missed.

The authors acknowledge Mirela Puleva for providing the structures and binding energies from the QUID dataset\cite{QUIDrep}. 
The research reviewed in this article was financially supported by different grants.
MB acknowledges financial support from the Luxembourg National Research Fund (INTER/DFG/18/12944860).
JACM acknowledges financial support from the Luxembourg National Research Fund (AFR PhD/19/MS, 13590856).
AT acknowledges financial support from the Luxembourg National Research Fund (C23/MS/18093472/MBD-in-BMD) and from the European Research Council (ERC, FITMOL, U-AGR-814900-C).
The calculations have been executed with the support of various computing facilities and with the support of different HPC grants.
In particular, some calculations have been carried out using the HPC facilities of the University of Luxembourg~\cite{VBCG_HPCS14} {\small (see \href{http://hpc.uni.lu}{hpc.uni.lu})}.
An award of computer time was provided by the Innovative and Novel Computational Impact on Theory and Experiment (INCITE) program.
This research used resources from the Argonne Leadership Computing Facility, which is a DOE Office of Science User Facility supported under Contract DE-AC02-06CH11357.
Some simulations were performed on the Luxembourg national supercomputer MeluXina, and the authors gratefully acknowledge the LuxProvide teams for their expert support.
MB acknowledges computational resources also from the EuroHPC Benchmark Access Call (EHPC-BEN-2023B07-035). 
All molecular structures have been visualized and rendered using VMD 2.0\cite{VMD2.0}. Plots are generated using Gnuplot 6.0\cite{gnuplot}. 
\end{acknowledgements}


\section*{Author Declarations}

\hspace{-4mm}{\bf \sffamily Conflict of Interest}\\[-2mm]

\hspace{-4mm}The authors have no conflicts to disclose.\\

\hspace{-4mm}{\bf \sffamily Author Contributions}\\[-2mm]

{ \small 
\hspace{-4mm}\textbf{Matteo Barborini:} Conceptualization (lead); Data curation (lead); Formal analysis (lead); Investigation (lead); Methodology (lead); Project Administration (lead); Software (lead); Supervision (lead); Validation (lead); Visualization (lead); Writing – original draft (lead).
\textbf{Jorge Charry:} Data curation (equal); Formal analysis (equal); Investigation(equal); Software (equal); Validation (equal); Visualization (equal); Writing – original draft (supporting).
\textbf{Matej Ditte:} Data curation (equal); Formal analysis (equal); Investigation(equal); Software (equal); Validation (equal); Visualization (equal); Writing – original draft (supporting).
\textbf{Andronikos Leventis:} Software (supporting); Writing – original draft (supporting).
\textbf{Georgios Kafanas:} Software (supporting); Writing – original draft (supporting).
\textbf{Alexandre Tkatchenko:} Project Administration (supporting); Supervision (supporting); Funding Acquisition (lead); Resources (lead); Writing – original draft (supporting).
}

%

\end{document}